\documentclass[prb,showpacs,twocolumn,aps,superscriptaddress,a4paper]{revtex4-1}

\usepackage{color}
\usepackage{dcolumn}
\usepackage{amsmath}
\usepackage{amssymb}
\usepackage{mathtools}
\usepackage{graphicx}
\usepackage{latexsym}
\usepackage{amsfonts}
\usepackage{amssymb}
\usepackage{comment}
\DeclareGraphicsExtensions{.pdf,.gif,.jpg}

\newcommand{\be}{\begin{equation}}  
\newcommand{\ee}{\end{equation}}
\newcommand{\beq}{\begin{eqnarray}}  
\newcommand{\eeq}{\end{eqnarray}}

\def\im{{\mathrm{i}}}
\def\ex{{\mathrm{e}}}

\let\chook\c

\def\bG{\mbox{\boldmath $G$}}  
\def\bH{\mbox{\boldmath $h$}}  
\def\bK{\mbox{\boldmath $K$}}

\def\bS{\mbox{\boldmath $S$}}

\def\unit{\mbox{\boldmath $1$}}

\def\bgG{\mbox{\boldmath $\varGamma$}}

\def\l{\lambda}

\def\F{\Phi}  
\def\c{\chi}  
\def\Q{\Psi}

\begin{document}


\title{Time-dependent Landauer-B{\"u}ttiker approach to charge pumping in AC-driven graphene nanoribbons }

\author{Michael Ridley}
\affiliation{The Raymond and Beverley Sackler Center for Computational Molecular and Materials Science, Tel Aviv University, Tel Aviv 6997801, Israel}

\author{Riku Tuovinen}
\affiliation{Max Planck Institute for the Structure and Dynamics of Matter, Center for Free Electron Laser Science, 22761 Hamburg, Germany}

\date{\today}  

\begin{abstract}
We apply the recently-developed partition-free time-dependent Landauer-B{\"u}ttiker (TD-LB) formalism to the study of periodically-driven transport in graphene nanoribbons (GNR). When an AC driving is applied, this formalism can be used to prove generic conditions for the existence of a non-zero DC component of the net current (pump current) through the molecular device. Time-reversal symmetry breaking in the driving field is investigated and found to be insufficient for a non-zero pump current. We then derive explicit formulas for the current response to a particular biharmonic bias. We calculate the pump current through different GNR configurations and find that the sign and existence of a non-zero pump current can be tuned by simple alterations to the static parameters of the TD bias. Further, we investigate transient currents in different GNR configurations. We find a selection rule of even and odd harmonic response signal depending on a broken dynamical inversion symmetry in the bias. 
\end{abstract}

  
\maketitle  


\section{Introduction}\label{sec:intro}

Active electronic circuit components can now be fabricated at the
nanoscale. These components typically consist of a molecular structure coupled to two or more conducting leads with an applied field that creates a net drift of electrons across the structure. The motivation for ever smaller transistors and wires lies in the available speed-up arising from both nanometre size
and THz intramolecular transport processes \cite{Ochoa2015}. Subsequent to the initial
proposal of molecular rectification \cite{Aviram1974}, chemical
fabrication techniques have lead to the realization of many interesting
devices, including single-electron transistors \cite{Rokhinson2000, Kafanov2009},
molecular wires \cite{Blum2003,Wohlgamuth2013}, frequency doublers
and detectors \cite{Swager1998,Iniguez-de-la-Torre2010} and switches
for fast memory storage \cite{JanvanderMolen2010,Liu2013}. 

The fabrication of molecular devices has caused
recent experimental work to increasingly focus on dynamical properties
of nanojunctions. The basic question of this field is to determine
the AC current response to an external periodic electromagnetic field
causing electronic excitations in the GHz-THz frequency range \cite{Burke2004,Li2004,Chaste2008,Lin2008,Zhang2014}.
A phenomenon known as photon-assisted tunneling (PAT), in which irradiated
tunnel junctions acquire additional peaks in their conductance spectra
has been experimentally demonstrated as an additional transport channel
in a variety of structures, beginning with tunneling between oxide
films in superconductors in 1962 \cite{Dayem1962}. 
These systems may find application in frequency-sensitive detectors \cite{Drexler1995,Blick1995}. In Ohmic
conductors, a bias that averages to zero over the driving signal time
period will always produce a zero average current. However, when the
current response to the applied bias is nonlinear, it is possible
for an external periodic field to have a vanishing integral over its
time period and still induce a directed current.
The mechanism behind this phenomenon is similar to one known in the
engineering literature as AC-DC conversion, or \emph{rectification}.
In the quantum transport literature, nanoscale structures are known
as quantum pumps if they possess periodically-varying parameters that
result in a net directed current, and the phenomenon of AC-DC conversion
itself is known as charge pumping \cite{Thouless1983,Niu1990,Brouwer2001}.
Quantum pumps can be created by the breaking of dynamical
symmetries in the driving bias \cite{Kohler2005,Denisov2007,Denisov2014},
in addition to adiabatically varying the physical parameters of the
nanojunction \cite{Brouwer2001,Yuge2012}.
Experimental demonstrations of two-parameter charge pumping have been
carried out \cite{Switkes1999,DiCarlo2003,Blumenthal2007}. Recent
theoretical work has demonstrated the possibility of single-parameter
charge pumping, in which a single periodic source may be used to generate
a pump current \cite{Vavilov2000,Moskalets2002,Arrachea2005,FoaTorres2005a,Agarwal2007}.
These predictions were confirmed experimentally for the low-GHz frequency
range in nanowires etched into semiconductor heterostructures \cite{Kaestner2008,Fujiwara2008,Kaestner2009,Kaestner2011}.

When modelling electron transport through molecular devices in response to an external time-dependent field, a trade-off is always made between accuracy and the computational time involved \cite{Gaury2014}. This trade-off becomes increasingly important as one moves from quantum dots to molecules with a more complex electronic structure. 

In principle, calculations of time-dependent electron transport should take into account the many-body nature of electron-electron and electron-phonon interactions \cite{svlbook}. However, in many systems of technological interest, it is possible to neglect explicit Coulombic interactions and work within an effectively ballistic transport regime. The most popular framework for the study of ballistic transport has been the Landauer-B\"{u}ttiker formalism, which constructs transport quantities from scattering state solutions of electron wave packets propagating through the molecular device \cite{Landauer1970,Buttiker1986}. This was initially a steady-state theory, but it has been generalised in recent years to describe systems driven by periodically-varying fields \cite{Moskalets2008} and voltage pulses \cite{Maciejko2006}. 

The present authors have contributed to a \emph{partition-free} time-dependent Landauer-B{\"u}ttiker (TD-LB) approach based on the nonequilibrium Green's function (NEGF) formalism~\cite{Tuovinen2013,Tuovinen2014,Ridley2015,Ridley2017}. In a partition-free transport setup the subsystems are prepared in equilibrium with each other before a bias is added to the electronic energies in the leads~\cite{Stefanucci2004}. This is in contrast to the so-called \emph{partitioned} switch-on approach~\cite{caroli1,caroli2,Jauho1994} where the subsystems are suddenly brought into contact. For ballistic transport the two approaches always give the same values of the current in the limit of long-times after the switch-on~\cite{Stefanucci2004}.

The TD-LB method neglects the electronic structure of the leads in the so-called wide-band limit approximation (WBLA), thus enabling an exact solution of the Kadanoff-Baym equations for all Green's functions of the molecular region. Initially, the partition-free approach was developed for constant bias switch-on processes \cite{Tuovinen2013,Tuovinen2014}, before it was generalised to the case of an arbitrarily time-dependent driving bias \cite{Ridley2015,Ridley2017}. In the latter work, solutions were given for all Green's functions of the molecular region in the two-time plane. We note that in the transient regime following the voltage quench, the partitioned and partition-free approaches yield a different current. However, recent work~\cite{Odashima2017} indicates that within the WBLA it is always possible to simulate the results of a partition-free switch-on within the partitioned approach of~\citet{Jauho1994}.

A popular method for treating periodically driven quantum systems is the Floquet scattering matrix approach~\cite{Kohler2005,Moskalets2002}. An explicit relation between the NEGF and Floquet formalism excluding many-body interactions has been found in Ref.~\onlinecite{Arrachea2006}. This relation has been used to formulate and efficiently solve the equations of motion for the Green's function for both charge and energy transport in ac-driven quantum systems~\cite{Wu2008,Arrachea2005,ARRACHEA2007,Ludovico2014}, and was also very recently applied to a similar graphene setup as in the present study~\cite{1709.00090}. We emphasise that, in the formalism of the present paper, no assumptions need to be made about the periodicity of the bias voltage. The Floquet formalism focuses on long-time dynamics, but increasingly studies have focused on the short-time transient current response to a sudden quench described by a change in the Hamiltonian of the system at some special `switch-on' time $t_{0}$. This change could involve a partitioned or a partition-free switch-on process discussed above. 

Within noninteracting models, several numerical time-propagation schemes have been developed that go beyond the WBLA. However, these methods scale with the number of time steps $N_{t}$ (to different powers)~\cite{Zhu2005} and can therefore become very expensive for large molecular structures. By contrast, the TD-LB formalism used in this paper scales with the system size as $N_{s}^{2}$, but 
does not scale at all with $N_{t}$, since all transport quantities are expressed as exact `single-shot' functions of the time. This enables a trivial parallelization of the calculation of the time-dependent current, as the different `single-shot' calculations can be distributed over the machine cores with no need to share information between them~\cite{Tuovinen2016thesis}.
So far, this formalism has been successfully applied to calculations of currents and populations in superconducting junctions~\cite{Tuovinen2016b}, molecular wires~\cite{Ridley2016a,Ridley2016b}, graphene nanoribbon (GNR)~\cite{Tuovinen2014,daRocha2015} and double quantum dot~\cite{Fukadai2017} systems for a variety of different time dependent biases. Recently, this formalism was extended to calculations of the transient current noise in extended molecular structures, for which the two-time Green's functions are essential~\cite{Ridley2017}.

Even though \emph{stationary} charge pumping in graphene based systems has been studied both experimentally~\cite{Connolly2013,Evelt2017} and theoretically~\cite{Prada2009,Prada2011,Torres2011,Ingaramo2013,Abdollahibour2014}, the \emph{transient regime}~\cite{Jnawali2013,Zhang2013} has remained fairly unexplored. In the present work, we will therefore apply the TD-LB formalism to the study of charge pumping in GNRs driven by periodic fields in both the transient and long time regimes. In Section \ref{sec:NEGF}, the TD-LB theory we will use is outlined, and then in Section \ref{sec:pump} it is applied to the problem of charge pumping in a generic molecular junction. We then outline the conditions needed to generate a finite net current across the system purely by tuning parameters of the driving bias. This set of conditions is referred to as the quantum pump symmetry theorem. Moving on to the actual implementation of the theory in Section \ref{sec:results}, we choose an appropriate bias to investigate the dependence of the long-time pumped current on dynamical symmetry breaking in both zigzag (zGNR) and armchair (aGNR) nanoribbon configurations. Following this, we present for these configurations fully time-dependent calculations of the current in response to a simple sinusoidal driving field.

\section{Theoretical background}\label{sec:theory}
\subsection{Hamiltonian and NEGF}\label{sec:NEGF}

The theory of NEGF is concerned with the calculation of ensemble averages in response to a bias switch-on event at time $t_{0}$. These ensemble averages are propagated along the so-called Konstantinov-Perel' (KP) contour $\gamma$, which contains two horizontal branches $C_{-}$, $C_{+}$ running from $t_{0}$ to $t$ and $t$ to $t_{0}$, respectively. It also includes a vertical branch $C_{M}$ running in the imaginary time direction from $t_{0}$ to $t_{0}-\im\beta$, where $\beta$ is the inverse temperature of the system. The latter `propagation' in imaginary time is mathematically isomorphic to a thermal averaging. In what follows, the variable $z$ is used to refer to generic contour times, and it is therefore necessary to specify a generic Hamiltonian for all values that $z$ takes along the KP contour:
\begin{equation}
\begin{split}
\hat{H}\left(z\right)=\underset{k\alpha\sigma}{\sum}\varepsilon_{k\alpha}\left(z\right)\hat{d}_{k\alpha\sigma}^{\dagger}\hat{d}_{k\alpha\sigma}+\underset{mn\sigma}{\sum}h_{mn}\left(z\right)\hat{d}_{m\sigma}^{\dagger}\hat{d}_{n\sigma} \\
+\underset{m,k\alpha\sigma}{\sum}\left[T_{mk\alpha}\left(z\right)\hat{d}_{m\sigma}^{\dagger}\hat{d}_{k\alpha\sigma}+T_{k\alpha m}\left(z\right)\hat{d}_{k\alpha\sigma}^{\dagger}\hat{d}_{m\sigma}\right] . \label{eq:Hamiltonian}
\end{split}
\end{equation}
The first term in this expression corresponds to the Hamiltonians
describing the individual reservoirs/leads, where the index $\alpha$ is used to label leads, and $k$
labels the $k$-th lead eigenstate. The second term corresponds
to the Hamiltonian of the molecular structure with indices $n$ and $m$
labeling molecular eigenstates. The final term describes the
coupling of the leads and the central system with the corresponding
matrix elements $T_{m,k\alpha}$, and $\sigma$ denotes the spin degree
of freedom of the electrons. The objects $\hat{d}_{k\alpha\sigma}$,
$\hat{d}_{m\sigma}$ and $\hat{d}_{k\alpha\sigma}^{\dagger}$, $\hat{d}_{m\sigma}^{\dagger}$
are annihilation and creation operators of the leads and the central
system.
 
The $i,j$-th component of
the one-particle Green's function on the KP contour is defined on the molecular basis as follows:
\begin{equation}
G_{ij}\left(z_{1},z_{2}\right)=-\im\frac{\mbox{Tr}\left[\ex^{-\beta\hat{H}^{\text{M}}}\hat{T}_{\gamma}\left[\hat{d}_{i,\text{H}}\left(z_{1}\right)\hat{d}_{j,\text{H}}^{\dagger}\left(z_{2}\right)\right]\right]}{\text{Tr}\left[\ex^{-\beta\hat{H}^{\text{M}}}\right]} .
\end{equation}
In this expression, the contour arguments $z_1$ and $z_2$ can be chosen to lie anywhere on $\gamma$, corresponding to a thermal average of pairs of creation/annihilation processes occurring both in and out of equilibrium. $\hat{T}_{\gamma}$ orders operators (Heisenberg picture) later on the KP contour to the left, and $\hat{H}^{\text{M}}=\hat{H}-\mu \hat{N}$ is the Matsubara hamiltonian describing the equilibrium system. Various components of the Green's function with useful physical meanings are defined by specifying the contour times, for example the lesser Green's function $G^{<}(t_1,t_2)$ may be obtained by choosing $z_1 \in C_{-}$ and $z_2 \in C_{+}$. 
 
To describe the switch-on of a bias within a partition-free approach, we assume that the lead-molecule coupling terms $T_{m,k\alpha}$ are present for the system both in and out of equilibrium, i.e. they are present for all values of $z$ on the KP contour.  In the time-dependent Hamiltonian previously studied within the TD-LB formalism, we assumed that the lead energies $\varepsilon_{k\alpha}\left(z\right)$ are contained in the following `block' of the hamiltonian energy matrix:
\begin{equation}
\left[\bH_{\alpha\alpha}\left(z\right)\right]_{kk'}=\left\{ \begin{array}{c}
\left(\varepsilon_{k\alpha}+V_{\alpha}\left(t\right)\right)\delta_{kk'},\; z\equiv t\in C_{-}\oplus C_{+}\\
\left(\varepsilon_{k\alpha}-\mu\right)\delta_{kk'},\; z\in C_{M} .
\end{array}\right.\label{eq:H_for_leads}
\end{equation}
Here, $V_{\alpha}\left(t\right)$, is an external bias applied to the leads of the nano junction. Finally, the energies of the molecular structure Hamiltonian are allowed to be shifted by a spatially homogeneous time-dependent field $V_{C}$ and a spatially local time independent term $u_{mn}$: 
\begin{equation}
\left[\bH_{CC}\left(z\right)\right]_{mn}=\left\{ \begin{array}{c}
h_{mn}+u_{mn}+V_{C}\left(t\right)\delta_{mn},\; z\in C_{-}\oplus C_{+}\\
h_{mn}-\mu\delta_{mn},\; z\in C_{M} .
\end{array}\right.\label{eq:H_for_central}
\end{equation}

In what follows, we will set $u_{mn}=0$ and $V_{C}\left(t\right)=0$, as the effects of the latter kind of time-dependence can be equivalently studied with a lead-independent term in the bias $V_{\alpha}\left(t\right)$ \cite{Ridley2016b}. In this case, one can re-express the Kadanoff-Baym equations for the different Green's function components in terms of an effective, non-Hermitian hamiltonian~\cite{svlbook}
\begin{equation}
\bH_{CC}^{\text{eff}}\equiv\bH_{CC}-\frac{\im}{2}\underset{\alpha}{\sum}\bgG_{\alpha}
\end{equation}
whose eigenvalues correspond to unstable eigenmodes of the molecular structure. These eigenmodes have a finite lifetime due to the presence of the level width matrix, which is defined in terms of the lead-molecule coupling
\begin{equation}
[\bgG_{\alpha}]_{mn}\left(\omega\right)=2\pi\sum_{k}T_{m,k\alpha}\delta\left(\omega-\varepsilon_{k\alpha}\right)T_{k\alpha,n}
\end{equation}
and assumed to be energy-independent in accordance with the WBLA, i.e. it is evaluated at the Fermi energy of lead $\alpha$. In terms of this effective hamiltonian, we may then derive the following exact expression for the greater/lesser Green's functions \cite{Ridley2016a}:
\beq
& & \bG_{CC}^{\gtrless}\left(t_{1},t_{2}\right) \nonumber \\
& = &\mp \im\int\frac{d\omega}{2\pi}f\left(\mp\left(\omega-\mu\right)\right)\underset{\alpha}{\sum}\bS_{\alpha}\left(t_{1},t_{0};\omega\right)\bgG_{\alpha}\bS_{\alpha}^{\dagger}\left(t_{2},t_{0};\omega\right) \nonumber \\\label{eq:greater/lesserGF}
\eeq
where we have defined the matrices $\bS_{\alpha}$ as follows
\begin{equation}
\bS_{\alpha}\left(t,t_{0};\omega\right)\equiv \ex^{-\im\bH_{CC}^{\text{eff}}\left(t-t_{0}\right)}\left[\bG_{CC}^{\text{r}}\left(\omega\right)-\im\bK_{\alpha}\left(t,t_{0};\omega\right)\right] . \label{eq:GFgreatless}
\end{equation}
In this expression, we have defined the frequency-dependent retarded Green's function
\begin{equation}
\bG_{CC}^{\text{r}}\left(\omega\right)=\left(\omega\unit_{CC} - \bH_{CC}^{\text{eff}} \right)^{-1}
\end{equation}
whose hermitian conjugate is the advanced component $\bG_{CC}^{\text{a}}\left(\omega\right)$~\cite{FetterWalecka,Danielewicz,svlbook}. In addition, we introduce the matrix object
 \begin{equation}
\bK_{\alpha}\left(t,t_{0};\omega\right)=\int_{t_{0}}^{t}d\bar{t}\ex^{-\im\left(\omega\unit_{CC}-\bH_{CC}^{\text{eff}}\right)\left(\bar{t}-t_{0}\right)}\ex^{-\im\psi_{\alpha}\left(\bar{t},t_{0}\right)} . \label{eq:K-integral}
\end{equation}
Here, the time-dependent bias $V_{\alpha}\left(t\right)$
of lead $\alpha$ enters into Eq. (\ref{eq:GFgreatless}) only via the phase factors 
\begin{equation}
\psi_{\alpha}\left(t_{1},t_{2}\right)\equiv\int_{t_2}^{t_1} d\tau\, V_{\alpha}\left(\tau\right)\label{eq:phasealpha}
\end{equation}
in the integrand of the $\bK_{\alpha}$ matrix. The time integrals in Eq.~(\ref{eq:GFgreatless}) can therefore often be removed analytically. The current in lead $\alpha$ is defined in terms of a particle number derivative, 
\begin{equation}
I_{\alpha}\left(t\right)\equiv q\left\langle \frac{d\hat{N}_{\alpha}\left(t\right)}{dt}\right\rangle 
\end{equation}
where $q$ is the electron charge and a factor of $2$ is included in the particle number to account for spin degeneracy. From this definition, one can derive the following rather compact expression for the current (with the convention $q=-1$) \cite{Ridley2015, Ridley2016a}:\begin{widetext}
\begin{equation}
I_{\alpha}\left(t\right)=\frac{1}{\pi}\int d\omega f\left(\omega-\mu\right)\,\mbox{Tr}_{C}\left[2\mbox{Re}\left[\im\ex^{\im\omega\left(t-t_{0}\right)}\ex^{\im\psi_{\alpha}\left(t,t_{0}\right)}\bS_{\alpha}\left(t,t_{0};\omega\right)\bgG_{\alpha}\right]-\bgG_{\alpha}\underset{\beta}{\sum}\bS_{\beta}\left(t,t_{0};\omega\right)\bgG_{\beta}\bS_{\beta}^{\dagger}\left(t,t_{0};\omega\right)\right] . \label{eq: CURRENT}
\end{equation}
\end{widetext}
We note in passing that whereas both time arguments in Eq. (\ref{eq:GFgreatless}) are needed to calculate current correlations \cite{Ridley2017}, for the first moment of the current only the single-time lesser Green's function is needed, corresponding to the second term in Eq. (\ref{eq: CURRENT}). 

\subsection{Symmetry conditions on a pump current}\label{sec:pump}
It is sometimes possible to induce a nonequilibrium process
that breaks the spatial symmetry of current flow by introducing a
term in the Hamiltonian that breaks time-reversal (TR) symmetry. Choosing
$\varOmega_{\alpha}=\varOmega_{D}=\varOmega_{\beta}$ to be the fundamental
driving frequency of the periodic signal in the leads, we now define
the pump current (also referred to as the DC component of the current
in the literature \cite{Arrachea2005,Kohler2005}) at time $\tau$
after the switch-on time:
\begin{equation}
I_{\alpha\beta}^{\left(\text{pump}\right)}\left(\tau\right)\equiv\frac{\varOmega_{D}}{2\pi}\int_{\tau}^{\tau+\frac{2\pi}{\varOmega_{D}}}dt\left(I_{\alpha}\left(t\right)-I_{\beta}\left(t\right)\right) . \label{eq:pump}
\end{equation}
There may be conditions under which a finite pumping current exists
in the transient regime following the switch-on, but not in the steady-state
limit, when transient modes in the current have decayed to zero. We
would like to define general conditions under which this is \emph{not}
true, i.e. we wish to investigate \emph{stable}
quantum pumps generated by dynamical driving fields satisfying the
following condition:
\begin{equation}
\underset{t_{0}\rightarrow-\infty}{\lim}I_{\alpha\beta}^{\left(\text{pump}\right)}\left(\tau\right) \neq 0 . \label{eq:stablepump}
\end{equation}
We work in the long time limit, because in this limit the charge pumped
per cycle should have settled into a steady-state value. In this limit,
it was shown in Ref.~\onlinecite{Ridley2015} that the long time current
can be expressed as\begin{widetext}
\beq
\underset{t_{0}\rightarrow-\infty}{\lim}I_{\alpha}(t) & = & \underset{t_{0}\rightarrow-\infty}{\lim}\frac{1}{\pi}\int d\omega f\left(\omega-\mu\right) \mbox{Tr}_{C}\left\{ 2\mbox{Re}\bgG_{\alpha}\ex^{\im\left(\omega\unit_{CC}-\bH_{CC}^{\text{eff}}\right)\left(t-t_{0}\right)}\ex^{\im\psi_{\alpha}\left(t,t_{0}\right)}\bK_{\alpha}\left(t,t_{0};\omega\right) \right.\nonumber\\
& & \hspace{50pt}\left. - \ \ex^{-\im\bH_{CC}^{\text{eff}}\left(t-t_{0}\right)}\bgG_{\alpha}\underset{\gamma}{\sum}\bK_{\gamma}\left(t,t_{0};\omega\right)\bgG_{\gamma}\bK_{\gamma}^{\dagger}\left(t,t_{0};\omega\right)\ex^{\im\left(\bH_{CC}^{\text{eff}}\right)^{\dagger}\left(t-t_{0}\right)}\right\} . \label{eq:asymptote-1}
\eeq
\end{widetext}
We now impose that the bias
driving the system is periodic with a basic driving frequency of $\varOmega_{D}$,
\begin{equation}
V_{\alpha}\left(t\right)=V_{\alpha}\left(t+\frac{2\pi}{\varOmega_{D}}\right) . \label{eq:Vbias}
\end{equation}
We can then show that the phase factor of the bias is periodic, and
can therefore be represented as a Fourier series with undetermined
coefficients:
\begin{eqnarray}
\ex^{-\im\psi_{\alpha}\left(t,-\infty\right)} & = & \exp\left(-\im\int_{-\infty}^{t-\frac{2\pi}{\varOmega_{D}}}d\tau V_{\alpha}\left(\tau\right)\right) \nonumber\\
 & = & \underset{n}{\sum}c_{n}^{\alpha}\ex^{-\im n\varOmega_{D}t} . \label{eq:FS1}
\end{eqnarray}
Here, we introduce the Fourier coefficient:
\begin{equation}
c_{n}^{\alpha}=\frac{\varOmega_{D}}{2\pi}\int_0^{\frac{2\pi}{\varOmega_{D}}}dt\ex^{-\im\psi_{\alpha}\left(t,-\infty\right)}\ex^{\im n\varOmega_{D}t} . \label{eq:FourierCoefficients}
\end{equation}
If, in addition, the bias is TR-symmetric, $V_{\alpha}\left(t\right)=V_{\alpha}\left(-t\right)$,
then the phase factor satisfies another useful property in the long-time
regime:
\begin{eqnarray}
\ex^{-\im\psi_{\alpha}\left(t,-\infty\right)} & = &\exp\left(\im\int_{-\infty}^{-t}d\tau V_{\alpha}\left(\tau\right)\right)\exp\left(\im\int_{\infty}^{-\infty}d\tau V_{\alpha}\left(\tau\right)\right)\nonumber \\
 & = & \left[\ex^{-\im\psi_{\alpha}\left(-t,-\infty\right)}\right]^{*} . 
\end{eqnarray}
Thus, from the identity~(\ref{eq:FS1}) we conclude that the Fourier
coefficients must satisfy the property 
\begin{equation}
c_{n}^{\alpha}=c_{n}^{\alpha*}\in\mathbb{R} . \label{eq:realcoeff}
\end{equation}
Thus, any bias which breaks TR symmetry leads to a phase with complex
Fourier coefficients $c_{n}^{\alpha}=\left|c_{n}^{\alpha}\right|\ex^{\im\phi_{\alpha}^{\left(n\right)}}$.

In Appendix~\ref{app:pumpsymm}, we use this property along with Eq.~(\ref{eq:FS1})
to prove a general result, which we refer to as the \emph{quantum
pump symmetry theorem}:

\emph{If (i) $V_{\alpha}\left(t\right)$ and/or $V_{\beta}\left(t\right)$
is given by a sum of more than one harmonic with frequencies that
are all integer multiples of $\varOmega_{D}$, (ii) in at least one of
the leads, TR symmetry is broken in at least one of the harmonics,
and (iii) the voltages satisfy $V_{\alpha}\left(t\right)\neq V_{\beta}\left(t\right)$,
then there is a non-zero net pump current running between the $\alpha$
and $\beta$ leads.}

Out of conditions (i)-(iii), condition (iii) of
this theorem is the only \emph{necessary}
condition for a pump current to exist, as there can be no pump current
if the bias in each lead is identical. The additional satisfaction
of conditions (i) and (ii) together with (iii) constitutes a \emph{sufficient},
but \emph{unnecessary} condition
for the existence of a non-zero pump current, i.e. the stronger statement
that the existence of a non-zero pump current requires (i) and (ii)
to hold in addition to (iii) is not true. A pump current could exist,
for instance, if TR symmetry was broken in one of the leads without
the additional assumption that the relation~(\ref{eq:phaserelate})
holds, which would be true if, e.g. the amplitude of the driving signal
was different in each lead. This quantum pump symmetry theorem however
gives experimentalists a means of generating a net current per driving
cycle with zero net bias per cycle and with no difference in the amplitude
of driving signals across the terminals of a nanodevice. It should
also be noted that in the so-called quantum ratchet effect, spatial
asymmetry of the junction in addition to a periodic driving is used
to generate a pumped current~\cite{Denisov2007}, but according to
the quantum pump symmetry theorem proven here, the system may be completely
spatially symmetric, so that $\bgG_{\alpha}=\bgG_{\beta}$,
and still there will be a reliable rectified current if the purely
dynamical conditions (i)-(iii) of this theorem are satisfied. Calculations
of the pump current through such a spatially symmetric system will
be presented in Section~\ref{sec:results}.

\section{Results}\label{sec:results}
\subsection{Transport setup}\label{sec:setup}
\begin{figure}[t]
\centering
\includegraphics[width=0.5\textwidth]{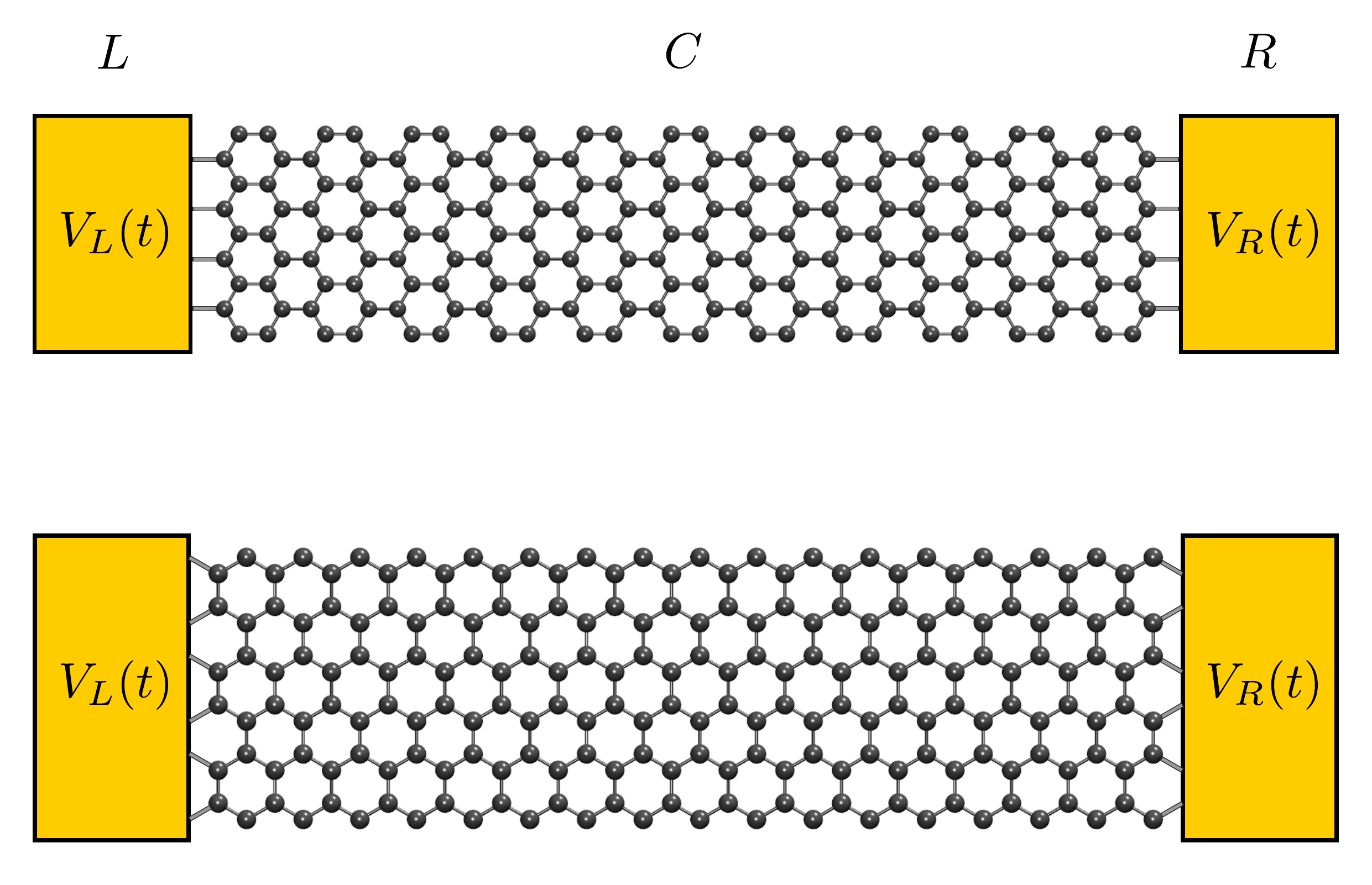}
\caption{Graphene nanoribbons studied in the ac transport simulations. Top: armchair edges longitudinally (transport direction), bottom: zigzag edges longitudinally.}
\label{fig:structures}
\end{figure}

We investigate ac transport in two representative graphene structures, see Fig.~\ref{fig:structures}. In the transport setup the central devices ($C$) are GNRs and they are similar in length ($\sim $ 4~nm), width ($\sim$ 1~nm) and in the number of carbon atoms ($\sim$ 200). The difference is in the orientation of the GNR, one having armchair edges in the transport direction while the other possesses zigzag edges. The left-most atoms are connected to the left lead ($L$) whereas the right-most atoms are connected to the right lead ($R$).
The structures are modelled in the tight-binding framework with a nearest-neighbour hopping $t_C = 2{.}7$~eV. From now on, we express all the energies in units of $t_C$. We wish to keep the tight-binding energy spectrum electron--hole symmetric and we only include the first nearest-neighbours; however, including second and third nearest-neighbours could be done with the same computational cost~\cite{Harju}. As stressed in the previous section, we work in the partition-free setup where the central region is being initially coupled to the lead environment. The global equilibrium temperature is set with $(k_{\text{B}}T)^{-1} = 100/t_C$, and with the symmetric energy spectrum we also set the chemical potential at $\mu=0$. The couping strength between the central region and the leads is included in the resonance width $\varGamma = \varGamma_{{L}} + \varGamma_{{R}} = t_C/40$ which corresponds to a weak coupling regime where WBLA is a good approximation~\cite{Verzijl2013}.

\subsection{Nonzero pump current with zero net driving}\label{sec:pumpresults}
We now wish to simulate the general result shown in Sec.~\ref{sec:theory}. We take the periodic bias profile for lead $\alpha$ to be of the form
\be
V_\alpha(t) = V_{\alpha} + A_\alpha^{(1)} \cos (p_1\varOmega_D t + \phi_\alpha) + A_\alpha^{(2)}\cos(p_2\varOmega_D t)\label{eq:biharmonicbias}
\ee
where $V_{\alpha}$ is a constant shift of the energy levels, $A_\alpha^{(1),(2)}$ are the amplitudes of the first and second harmonic driving, $\varOmega_D$ is the frequency of the driving, and $\phi_\alpha$ is the phase shift of the first harmonic. This profile is periodic in $T\equiv \varOmega_D/(2\pi)$ as in $V_\alpha(t+T) = V_\alpha(t)$. Formulas for the time-dependent current and pump current corresponding to this choice of bias are presented in Appendix~\ref{app:details}.

In Fig.~\ref{fig:pumpcurrent} we plot the left--right pump current $I_{\text{LR}}$ versus the phase difference $\phi \equiv |\phi_L-\phi_R|$ with varying bias strengths. To emphasize the dependence on $\phi$ alone, we choose the frequencies of the two harmonics in Eq. \eqref{eq:biharmonicbias} to be equal, $p_1=1=p_2$. We take the oscillation amplitude to be half of the constant bias shift, $A_\alpha^{(1)}  =  A_\alpha^{(2)} = V_\alpha/2$, and the oscillation frequency to be $\varOmega_D=t_C$. We clearly observe the condition for zero pump current when $\phi = 0,\pm2n\pi$ for integer $n$m and the maximal values are obtained when $\phi = \pm n\pi$. In addition, the curves are symmetric around $\phi=0$. This can be understood by choosing all parameters in Eq.~\eqref{eq:pumparbitrary} to be lead-independent with the exception of the phase, in which case one obtains the following pump current for the two-lead system:
\begin{widetext}
\beq
I_{LR}^{\left(\text{pump}\right)} & = & \frac{1}{\pi}\underset{j,k}{\sum}\underset{r,r',s,s'}{\sum}\delta_{ss}^{rr'}\left(p_{i}\right)J_{r}\left(\frac{A^{\left(1\right)}}{p_{1}\varOmega_{D}}\right)J_{r'}\left(\frac{A^{\left(1\right)}}{p_{1}\varOmega_{D}}\right)J_{s}\left(\frac{A^{\left(2\right)}}{p_{2}\varOmega_{D}}\right)J_{s'}\left(\frac{A^{\left(2\right)}}{p_{2}\varOmega_{D}}\right) \nonumber \\
& \times & \frac{\left\langle \varphi_{k}^{R}\right|\bgG_{L}\left|\varphi_{j}^{R}\right\rangle \left\langle \varphi_{j}^{L}\right|\bgG_{R}\left|\varphi_{k}^{L}\right\rangle +\left\langle \varphi_{k}^{R}\right|\bgG_{R}\left|\varphi_{j}^{R}\right\rangle \left\langle \varphi_{j}^{L}\right|\bgG_{L}\left|\varphi_{k}^{L}\right\rangle }{\left\langle \varphi_{j}^{L}\mid\varphi_{j}^{R}\right\rangle \left\langle \varphi_{k}^{R}\mid\varphi_{k}^{L}\right\rangle \left(\bar{\varepsilon}_{j}-\bar{\varepsilon}_{k}^{*}\right)}\left[\ex^{-\im\left(r-r'\right)\phi_{L}}-\ex^{-\im\left(r-r'\right)\phi_{R}}\right] \nonumber \\
& \times & \left[\Psi\left(\frac{1}{2}-\frac{\beta}{2\pi i}\left(\bar{\varepsilon}_{j}-\mu-V-\varOmega_{D}\left(p_{1}r+p_{2}s\right)\right)\right)-\Psi\left(\frac{1}{2}+\frac{\beta}{2\pi i}\left(\bar{\varepsilon}_{k}^{*}-\mu-V-\varOmega_{D}\left(p_{1}r'+p_{2}s'\right)\right)\right)\right] . \label{eq:pumpLRsymm}
\eeq
\end{widetext}
Note the similarity of this formula to the generic case in Eq. (\ref{eq:generalpump}),
and in particular note how the phase difference $\left[\ex^{-\im\left(r-r'\right)\phi_{L}}-\ex^{-\im\left(r-r'\right)\phi_{R}}\right]$
has an identical structure to Eq. (\ref{eq:amprelate}). The presence
of the function $\delta_{ss'}^{rr'}\left(p_{i}\right)$ guarantees
that there is no pump current without a second harmonic in any lead
(i.e. when $A_{\gamma}^{\left(2\right)}=0$ for all $\gamma$), as
in this case the summations over $s,s'$ vanish and $r=r'$. Also, we note that from the relation 
\be
\ex^{-\im\left(r-r'\right)\phi_{L}}-\ex^{-\im\left(r-r'\right)\phi_{R}}\propto-2\im\sin\left(\left(r-r'\right)\frac{\phi}{2}\right)
 \label{eq:phasediff}
\ee
it is evident that the pump current in this system must be zero when $\phi=0, \pm2n\pi$ for integer n, and that it possesses maximal values when $\phi_{LR}=\pm n\pi$. This behaviour differs qualitatively from the dependence on a phase difference $\delta$ between oscillating parameters in standard treatments of double parametric quantum pumping~\cite{Arrachea2005}. In these approaches the pumped current is usually proportional to $\sin(\delta)$ and is therefore asymmetric about its zero points, which occur at multiples of $\pm\pi$. As the formula for $I_{LR}^{\left(\text{pump}\right)}$ must be unchanged under the exchange of indices $r\leftrightarrow r'$, the expression \eqref{eq:phasediff} also implies that the pump current is unchanged upon reversal of sign $\phi\leftrightarrow-\phi$. Both these facts are reflected in the pump current characteristics of Fig.~\ref{fig:pumpcurrent}. 

\begin{figure}[t]
\centering
\includegraphics[height=0.15\textheight]{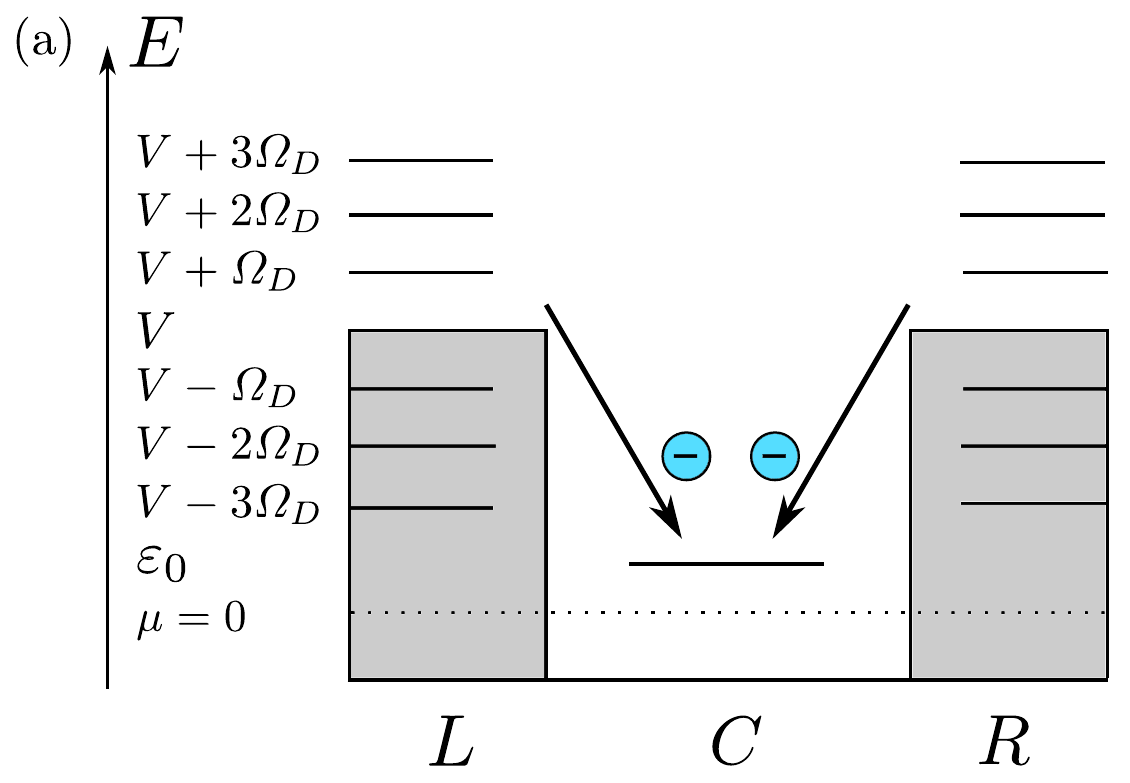}
\includegraphics[height=0.15\textheight]{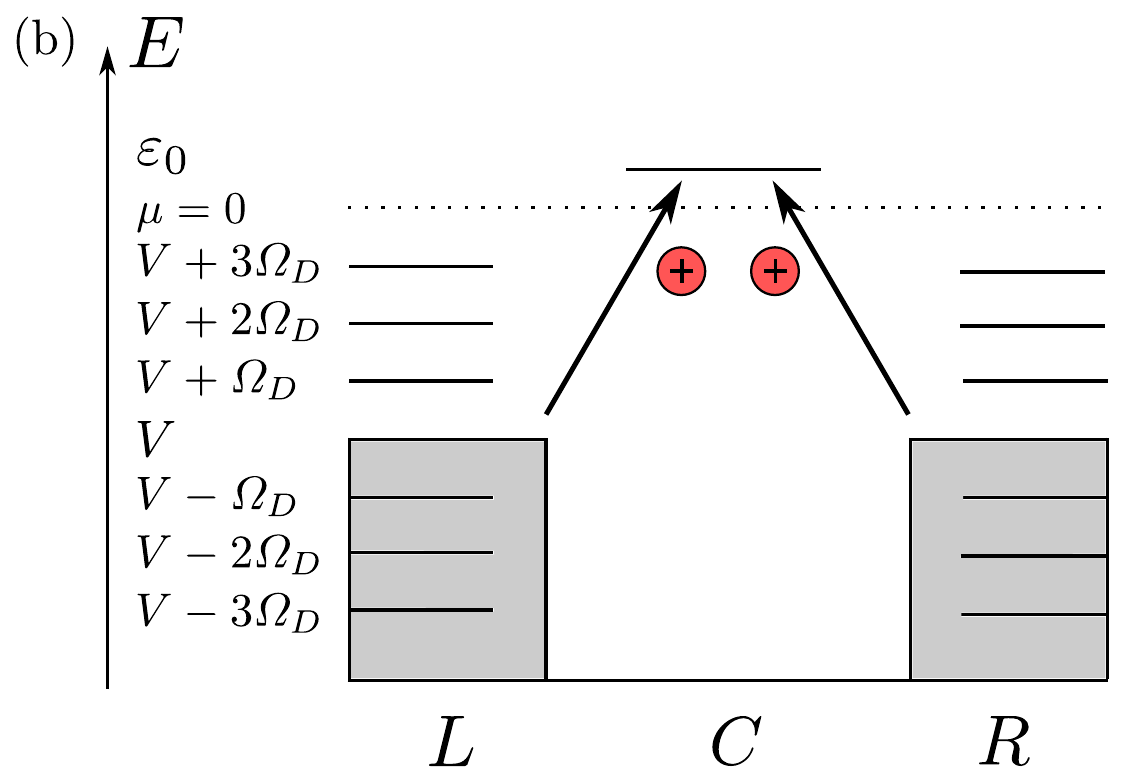}
\caption{Single-level schematic: electron transfer processes in the $V>\varepsilon_{0}$ case (a) are replaced by hole transfer processes in the $V<\varepsilon_{0}$ case (b) following inversion of the bias about the dot energy.}
\label{fig:particlehole}
\end{figure}

Interestingly, we also find that the pump current changes sign as the applied bias crosses the value $V_{\alpha}=t_C$. For single-level transport, i.e. when the molecular Hamiltonian is equal to the level energy, $\bH_{CC}^{\text{eff}}\equiv{\varepsilon}_{0}-\frac{\im}{2}\varGamma$, this sign reversal can be explained with a simple particle-hole symmetry argument~\cite{Ridley2017thesis}, illustrated schematically in Fig.~\ref{fig:particlehole} with $\varepsilon_{0}=1$. Due to the periodic driving in the leads, transport channels are opened up between the dot energy and various photon-assisted sidebands, which occur with a weighting given by some product of Bessel functions appearing in the summation of Eq. (\ref{eq:generalpump}). When $V>\varepsilon_{0}$, shown in Fig.~\ref{fig:particlehole}(a), electrons tunnel from the sidebands onto the level, crossing an energy gap of $\left|\mu+V+n\varOmega_{D}-\varepsilon_{0}\right|$. When $V<\varepsilon_{0}$, shown in Fig.~\ref{fig:particlehole}(b), electron tunnelling processes are replaced by corresponding oppositely-charged hole transfer processes with an energy gap of $-\left|\varepsilon_{0}+\mu-V-n\varOmega_{D}\right|$. These processes occur with the same weighting in the transmission as correspondent electron transfer processes in the $V>\varepsilon_{0}$ case, as can be seen by evaluating the particle/hole populations for each ordering of $V$ and $\varepsilon_{0}$~\cite{Ridley2017thesis}. This argument may be extended to the graphene structures studied in the present work, if we note that the density of states for each structure has a pronounced peak (a van Hove singularity) at $t_C$. The situation is almost as if we had only one dominant energy level at $E=t_C$, and when the bias strength is close in energy with the resonant level the pump current changes sign. We note that analogous bias-dependent current sign-switching about a point of symmetry in the GNR energy spectrum was observed experimentally in Ref.~\onlinecite{Connolly2013}.

\begin{figure}[t]
\centering
\includegraphics[width=0.5\textwidth]{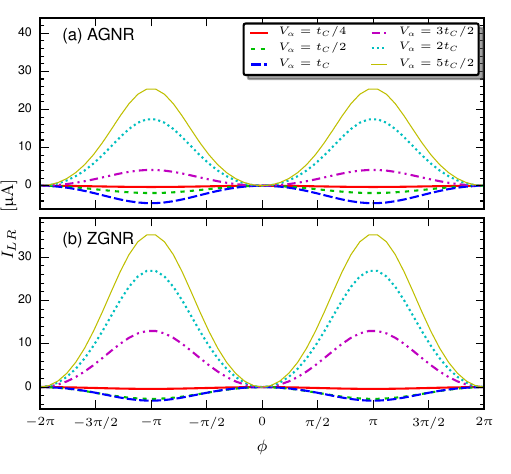}
\caption{Pump current versus the phase difference in armchair (a) and zigzag (b) graphene nanoribbons with varying bias strengths.}
\label{fig:pumpcurrent}
\end{figure}

\begin{figure}[t]
\centering
\includegraphics[width=0.5\textwidth]{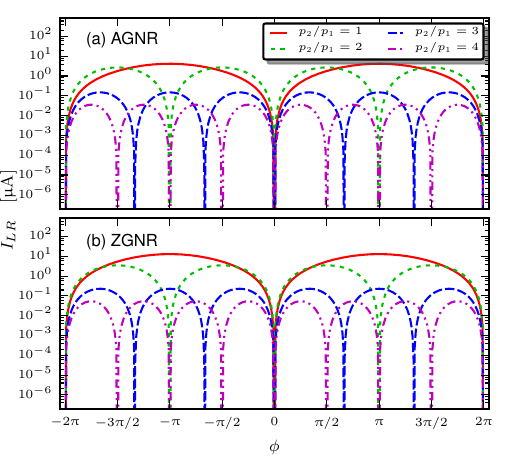}
\caption{Pump current versus the phase difference in armchair (a) and zigzag (b) graphene nanoribbons with different values of the frequency ratio $p_{2}/p_{1}$, where $p_{1}=1$}
\label{fig:pdependence}
\end{figure}

In Fig.~\ref{fig:pdependence} we display plots of the $\phi$-dependent pump current in the different GNRs, where the frequency ratio $p_{2}/p_{1}$ is allowed to vary. We choose the fixed bias shift $V_{\alpha}=3t_{C}/2$, with all other parameters as in Fig.~\ref{fig:pumpcurrent}. The current is plotted on a logarithmic scale as the presence of higher frequency modes causes suppression of high-order Bessel functions in the summation of Eq.~\eqref{eq:pumpLRsymm}. In all cases displayed, the second harmonic in Eq.~\eqref{eq:biharmonicbias} drives the system with a frequency that is a multiple of the frequency of the first harmonic. This causes additional nodes to form at $\pm n2\pi/p_{2}$ in the phase-dependent pump current, corresponding to `off' states of the electric signal. Formally, this is because the phase dependence is coupled to $p_{2}$ through the presence of the generalized Kronecker $\delta_{ss}^{rr'}\left(p_{i}\right)$ in Eq.~\eqref{eq:pumpLRsymm}. The definition of this object, Eq. \eqref{eq:Kronp1p2} implies that we can make the replacement $r-r'\rightarrow-p_{2}\left(s-s'\right)$ everywhere in Eq.~\eqref{eq:pumpLRsymm}, and therefore in the phase-dependent factor~\eqref{eq:phasediff}, so that the nodes of the pump current appear when $p_{2}\left(s'-s\right)\phi/2=\pm n\pi$. This complex switching behaviour provides experimentalists with a larger parameter space for the generation of `off' states in a GNR-based switch.

\begin{figure}[t]
\centering
\includegraphics[width=0.5\textwidth]{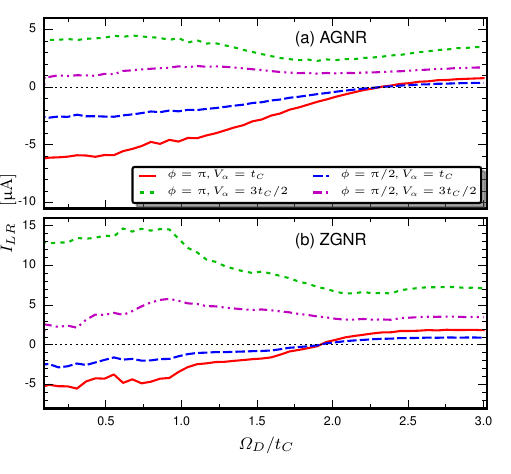}
\caption{Pump current versus the driving frequency in armchair (a) and zigzag (b) graphene nanoribbons with different bias and phase values.}
\label{fig:drive}
\end{figure}

In Fig.~\ref{fig:drive} we also show how the pump current through the GNRs at certain phase difference and bias strength depends on the driving frequency, for $\varOmega/t_C \geq 0.1$. Verifying the observation in Fig.~\ref{fig:pumpcurrent} also here we see how the curves with $V_{\alpha}=t_C$ and $V_{\alpha}=3t_C/2$ are of opposite sign. Interestingly, we also observe the pump current changing its sign as a function of the driving frequency. For instance, the pump current with $V_{\alpha}=t_C$ and $\phi=\pi/2,\pi$ becomes positive (for both armchair and zigzag geometries) with higher driving frequencies. The $\varOmega_{D}\to 0$ limit of the pump current is discussed in Appendix~\ref{app:adiabatic}, where it is shown to converge to zero at zero driving frequency. For different values of $\phi$, the regime for driving frequencies that are small but still finite ($\varOmega/t_C < 0.1$) may be explored by increasing the number of Bessel functions in the summations over $r,r',s,s'$ until convergence is achieved.

\subsection{High-harmonics response}
Now we investigate the full time-dependent response of the GNRs to two different ac bias voltages. In contrast to the previous pump current calculations, now the bias profiles consist of a monoharmonic drive: $V_{{L}}(t) = V_0 + A\cos(\varOmega_D t)$ and $V_{{R}}(t) = -V_{{L}}(t)$. In the following we consider the effects of odd inversion symmetry of the bias profile with respect to half the period: 
\be
V_\alpha(t+T/2) = -V_{\alpha}(t) . \label{eq:hpsymmetry}
\ee

\begin{figure}[t]
\centering
\includegraphics[width=0.5\textwidth]{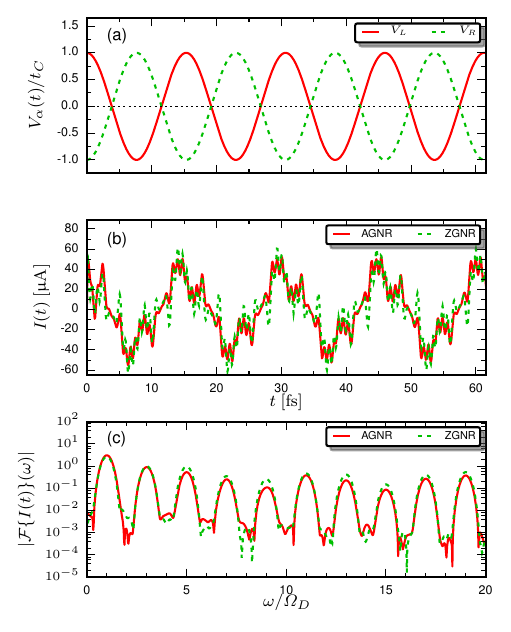}
\caption{Time-dependent response of the graphene nanoribbons to an odd-inversion-symmetric drive. (a) Time-dependent bias voltage profile, (b) transient current, (c) Fourier transform of the transient current (arbitrary units).}
\label{fig:tdcurrent_odd}
\end{figure}

By considering the Fourier series representation of this kind of bias, it is straightforward to show from the property~\eqref{eq:hpsymmetry} that all the even harmonics in the series vanish. In Fig.~\ref{fig:tdcurrent_odd}(a), we consider the current response to out-of phase biases applied to each lead with constant term $V_0=0$. This bias profile satisfies the odd inversion symmetry condition~\eqref{eq:hpsymmetry}. The amplitude and the frequency of the oscillation are set, respectively, to $A=t_C$, $\varOmega_D = t_C/10 = 2\pi/T$ with period $T\approx 15$~fs. In Fig.~\ref{fig:tdcurrent_odd}(b) we show the transient current, $I(t) \equiv (I_{{L}}(t)+I_{{R}}(t))/2$, through the different GNR structures with the bias voltage profile of Fig.~\ref{fig:tdcurrent_odd}(a). Since our unit of energy is $\epsilon=1$~eV, we have the conversions for the units of time and current by $t=\hbar/\epsilon \approx 6{.}58\times 10^{-16}$~s and $I=e\epsilon/\hbar\approx 2{.}43 \times 10^{-4}$~A, respectively. To study the response more in detail we take the absolute value of the Fourier transform (FT) of the time-dependent current signal, see Fig.~\ref{fig:tdcurrent_odd}(c). The FT is computed from an extended signal of the one shown in panel (b), and Blackman window filtering is used. The FT displays peaks at the odd harmonics of the basic driving frequency, $\omega = (2n+1)\varOmega_D$, up to very high-harmonic order. The appearance of odd harmonics only is due to the spectrum of the unbiased Hamiltonian being electron-hole symmetric and the odd inversion symmetry of the applied ac bias profile. Physically, one may consider the picture in Fig.~\ref{fig:particlehole} (a), with photon-assisted tunnelling of electrons only permitted to occur from sidebands lying at odd multiples of $\varOmega_D$, so that these are the resonant frequencies of the current. We note that the presence of these effects is indicative of a system operating far beyond the linear response regime.

\begin{figure}[t]
\centering
\includegraphics[width=0.5\textwidth]{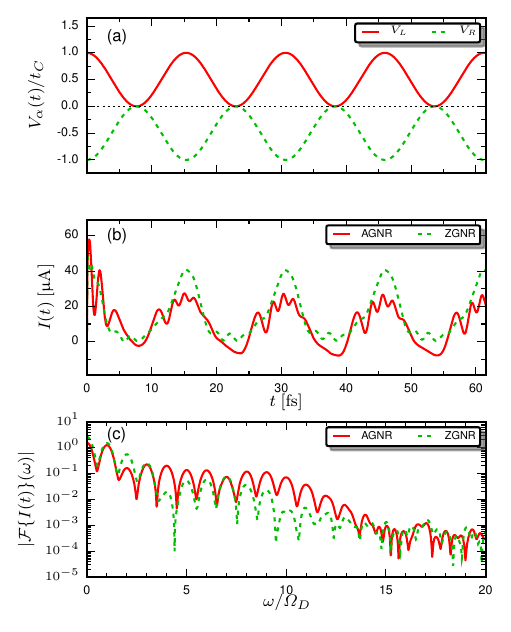}
\caption{Time-dependent response of the graphene nanoribbons to a broken-inversion-symmetric drive. (a) Time-dependent bias voltage profile, (b) transient current, (c) Fourier transform of the transient current (arbitrary units).}
\label{fig:tdcurrent_even}
\end{figure}

In the second case, we break the odd inversion symmetry of the applied bias with a constant shift term, see Fig.~\ref{fig:tdcurrent_even}(a). In this case we set $V_0=t_C/2$ and $A=t_C/2$ and the driving frequency is chosen to be the same as in the first case. In Fig.~\ref{fig:tdcurrent_even}(b) we show the transient behaviour of the current with the on-off bias voltage profile in Fig.~\ref{fig:tdcurrent_even}(a). Now, high harmonics of even order will develop since we broke the odd inversion symmetry of the applied bias voltage~\cite{Myohanen2010}, and there is a non-trivial contribution to the current from photon-assisted sidebands occurring at even multiples of the fundamental driving frequency. Therefore, in addition to the odd harmonics in the previous case, we observe peaks [see Fig.~\ref{fig:tdcurrent_even}(c)] at $\omega = 2n\varOmega_D$, also up to very high harmonics. The appearance of the odd-even harmonics could also be controlled, e.g. by breaking the particle-hole symmetry of the unbiased Hamiltonian via $2$nd and $3$rd nearest-neighbour hoppings~\cite{CastroNeto2009}.

In this bias voltage range, no qualitative differences are found between the transient currents in the armchair and zigzag geometries, albeit certain peaks in the Fourier spectrum might be a little more pronounced which is due to structure-specific lead--ribbon and intraribbon transitions: For instance, around $\omega = 10 \varOmega_D = t_C$ the overall intensity of the harmonics is enhanced due to mixing with the intraribbon HOMO--LUMO-like transitions. In both cases the time-dependent current [Figs.~\ref{fig:tdcurrent_odd}(b) and~\ref{fig:tdcurrent_even}(b)] shows a rapid transient oscillation relaxing to a periodic steady-state like solution in a timescale of about $10$~fs. After this, a time-varying persistent oscillation is observed with a superimposed ``ringing'' feature due to the internal structure of the GNRs~\cite{Jauho1994,Ridley2015}.

\section{Conclusions}\label{sec:concl}
The time-dependent Landauer--B{\"u}ttiker formalism was used to study charge pumping in a generic molecular junction, enabling fast simulation of the transient, non-adiabatic and steady-state electron transport regimes for the same computational cost. It was shown via the quantum pump symmetry theorem that a non-zero net current across a nanojunction can be generated purely by tuning the parameters of the driving bias.

As an application of this theory we studied graphene nanoribbons driven by periodic fields in both the transient and long time regimes. We illustrated the consequences of the quantum pump symmetry theorem by choosing an appropriate bias to investigate the dependence of the long-time pumped current on dynamical symmetry breaking in different GNR configurations. The results of these calculations show that discrete on/off states of the current can be generated from analog waves, which has implications for the design of graphene-based GHz-THz frequency modulators~\cite{Gao2014}, switches~\cite{Lin2010} and frequency detectors~\cite{Koppens2014}.

We also presented for these GNR configurations fully time-dependent calculations of the current in response to a simple sinusoidal driving field. By varying the inversion symmetry of the applied AC bias profiles, the current response \emph{selectively} generates very high-order odd or even harmonics of the basic driving frequency. This will prove useful in the design of graphene-based switches or frequency detectors. Future work will involve an extension of the method presented here to calculations of the time-dependent current noise of GNR and other 2D materials~\cite{Ridley2017}.

\acknowledgments
We would like to thank Angus MacKinnon for the suggestion of investigating quantum pumping with the TD-LB formalism. This work was financially supported by the Raymond and Beverly Sackler Center for Computational Molecular and Materials Science, Tel Aviv University (M.R.) and by the DFG (Grant No. SE 2558/2-1) through the Emmy Noether program (R.T.).

\appendix

\section{Details of the quantum pump symmetry theorem}\label{app:pumpsymm}
The bias in Eq.~\eqref{eq:Vbias} can always be represented as a sum of $N$ periodic
terms with frequencies that are integer multiples of $\varOmega_{D}$,
\begin{equation}
V_{\alpha}\left(t\right)=\underset{n=1}{\overset{N}{\sum}}V_{\alpha}^{\left(n\right)}\left(t\right)=\underset{n=1}{\overset{N}{\sum}}V_{\alpha}^{\left(n\right)}\left(t+\frac{2\pi}{p_{n}\varOmega_{D}}\right)\label{eq:VFS}
\end{equation}
where $p_{n}$ is an integer depending on the position
$n$ in the series. For example the Fourier Series of $V_{\alpha}\left(t\right)$
has the form of Eq. (\ref{eq:VFS}) with $N\rightarrow\infty$ and
with the $p_{n}$ ranging from minus to plus infinity. The trivial
case of $N=1$ and $p_{1}=1$ returns us to the basic periodic form
(\ref{eq:Vbias}), but it should be noted that whereas the full signal
$V_{\alpha}\left(t\right)$ has a time period of $\frac{2\pi}{\varOmega_{D}}$,
it does not have the periodicity $\frac{2\pi}{p_{n}\varOmega_{D}}$ of
its constituent terms. This allows us to study not just the periodicity
of the driving field, but also the decomposition of the driving field
into harmonics of the fundamental driving frequency and the effect
of the interplay of these harmonics on the pump current. With the
representation of Eq. (\ref{eq:VFS}), the exponential phase factor
is a product of periodic functions:
\beq
\ex^{-\im\psi_{\alpha}\left(t,-\infty\right)} & = & \exp\left(-\im\int_{-\infty}^t d\tau V_{\alpha}\left(\tau\right)\right) \nonumber \\
& = &\overset{N}{\underset{n=1}{\prod}}\exp\left(-\im\int_{-\infty}^t d\tau V_{\alpha}^{\left(n\right)}\left(\tau\right)\right)\nonumber \\
& = & \overset{N}{\underset{n=1}{\prod}}\exp\left(-\im\int_{-\infty}^{t-\frac{2\pi}{p_{n}\varOmega_{D}}}d\tau V_{\alpha}^{\left(n\right)}\left(\tau\right)\right) . \qquad
\label{eq:phaseINT}
\eeq
Each term in this product can therefore be expressed as a Fourier
Series, as in Eq. (\ref{eq:Vbias}):
\be
\ex^{-\im\psi_{\alpha}\left(t,-\infty\right)} = \sum_{\mathclap{m_{1}\ldots m_{N}}}c_{m_{1}}^{\alpha}\ldots c_{m_{N}}^{\alpha}\ex^{-\im\varOmega_{D}\left(p_{1}m_{1}+p_{2}m_{2}+\ldots+p_{N}m_{N}\right)t}\label{eq:FSphase}
\ee
where the $c_{m_{i}}^{\alpha}$ are undetermined lead-dependent Fourier
coefficients whose value depends upon the particular form of $V_{\alpha}^{\left(i\right)}\left(t\right)$. When we substitute Eq. (\ref{eq:FSphase}) into Eq. (\ref{eq:asymptote-1})
and evaluate the time integrals, one obtains:\begin{widetext}
\beq
\underset{t_{0}\rightarrow-\infty}{\lim}\left(I_{\alpha}(t)-I_{\beta}(t)\right) & = &\frac{1}{\pi}\int d\omega f\left(\omega-\mu\right)\sum_{\mathclap{\substack{m_{1},\ldots,m_{N}\\l_{1},\ldots,l_{N}}}} \ \mbox{Tr}_{C}\left[2\mbox{Re}\left[\im\left(\bgG_{\alpha}c_{m_{1}}^{\alpha}c_{l_{1}}^{\alpha*}\ldots c_{m_{N}}^{\alpha}c_{l_{N}}^{\alpha*}-\bgG_{\beta}c_{m_{1}}^{\beta}c_{l_{1}}^{\beta*}\ldots c_{m_{N}}^{\beta}c_{l_{N}}^{\beta*}\right)\right.\right.\nonumber \\
& & \hspace{35pt}\left.\times \ex^{-\im\varOmega_{D}\left(p_{1}\left(m_{1}-l_{1}\right)+\ldots+p_{N}\left(m_{N}-l_{N}\right)\right)t}\bG^{\text{r}}\left(\omega+\varOmega_{D}\left(p_{1}m_{1}+\ldots+p_{N}m_{N}\right)\right)\right]\nonumber \\
& & \hspace{35pt}-\left(\bgG_{\alpha}-\bgG_{\beta}\right)c_{m_{1}}^{\gamma}c_{l_{1}}^{\gamma*}\ldots c_{m_{N}}^{\gamma}c_{l_{N}}^{\gamma*}\ex^{-\im\varOmega_{D}\left(p_{1}\left(m_{1}-l_{1}\right)+\ldots+p_{N}\left(m_{N}-l_{N}\right)\right)t} \nonumber \\
& & \hspace{35pt}\left.\times\bG^{\text{r}}\left(\omega+\varOmega_{D}\left(p_{1}m_{1}+\ldots+p_{N}m_{N}\right)\right)\bgG_{\gamma}\bG^{\text{a}}\left(\omega+\varOmega_{D}\left(p_{1}l_{1}+\ldots+p_{N}l_{N}\right)\right)\right] . \label{eq:FScurrent}
\eeq
Note that the expression (\ref{eq:FScurrent}) is periodic with time
period $2\pi/\varOmega_{D}$, and therefore the long-time pump current
can be evaluated over the integration range $\left[0,2\pi/\varOmega_{D}\right]$:
\beq
I_{\alpha\beta}^{\left(\text{pump}\right)} & \equiv & \underset{t_{0}\rightarrow-\infty}{\lim}I_{\alpha\beta}^{\left(\text{pump}\right)}\left(\tau\right) = \frac{\varOmega_{D}}{2\pi}\int_0^{\frac{2\pi}{\varOmega_{D}}} dt\underset{t_{0}\rightarrow-\infty}{\lim}\left(I_{\alpha}\left(t\right)-I_{\beta}\left(t\right)\right) \nonumber \\
& = & \frac{1}{\pi}\int d\omega f\left(\omega-\mu\right)\sum_{\mathclap{\substack{m_{1},\ldots,m_{N}\\l_{1},\ldots,l_{N}}}} \ \delta_{m_{i}l_{i}}\left(N\right)\,\mbox{Tr}_{C}\left[2\mbox{Re}\left[\im\left(\bgG_{\alpha}c_{m_{1}}^{\alpha}c_{l_{1}}^{\alpha*}\ldots c_{m_{N}}^{\alpha}c_{l_{N}}^{\alpha*}-\bgG_{\beta}c_{m_{1}}^{\beta}c_{l_{1}}^{\beta*}\ldots c_{m_{N}}^{\beta}c_{l_{N}}^{\beta*}\right)\right.\right. \nonumber \\
& &\hspace{40pt}\left.\left.\times\bG^{\text{r}}\left(\omega+\varOmega_{D}\left(p_{1}m_{1}+\ldots+p_{N}m_{N}\right)\right)\right]\right] \nonumber \\
& & \hspace{40pt}-\left(\bgG_{\alpha}-\bgG_{\beta}\right)\underset{\gamma}{\sum}c_{m_{1}}^{\gamma}c_{l_{1}}^{\gamma*}\ldots c_{m_{N}}^{\gamma}c_{l_{N}}^{\gamma*} \nonumber \\
& & \hspace{40pt}\left.\times\bG^{\text{r}}\left(\omega+\varOmega_{D}\left(p_{1}m_{1}+\ldots+p_{N}m_{N}\right)\right)\bgG_{\gamma}\bG^{\text{a}}\left(\omega+\varOmega_{D}\left(p_{1}l_{1}+\ldots+p_{N}l_{N}\right)\right)\right] . \label{eq:generalpumpasymm}
\eeq
Here we make the definition:
\begin{equation}
\delta_{m_{i}l_{i}}\left(N\right)\equiv\begin{cases}
1, & p_{1}\left(m_{1}-l_{1}\right)+\ldots+p_{N}\left(m_{N}-l_{N}\right)=0\\
0, & \text{else}.
\end{cases}
\end{equation}
For instance, when $N=2$, this can be written
\be
\delta_{m_{i}l_{i}}\left(N\right) = \delta_{p_{1}m_{1},p_{1}l_{1}}\delta_{p_{2}m_{2},p_{2}l_{2}}+\delta_{p_{1}m_{1},p_{2}l_{2}}\delta_{p_{2}m_{2},p_{1}l_{1}} + \delta_{p_{1}m_{1},-p_{2}m_{2}}\delta_{p_{1}l_{1},-p_{2}l_{2}} . \label{eq:kroneckerexpand}
\ee
To simplify Eq. (\ref{eq:generalpumpasymm}), we use the identity
\be
\bG^{\text{r}}\left(\omega+A\right)-\bG^{\text{a}}\left(\omega+B\right) = \bG^{\text{r}}\left(\omega+A\right)\left(B-A-\im\bgG\right)\bG^{\text{a}}\left(\omega+B\right)\label{eq:gdiffID}
\ee
thereby obtaining
\beq
& & I_{\alpha\beta}^{\left(\text{pump}\right)}\nonumber \\
& = & \frac{1}{\pi}\int d\omega f\left(\omega-\mu\right)\sum_{\mathclap{\substack{m_{1},\ldots,m_{N}\\
l_{1},\ldots,l_{N}}}}\delta_{m_{i}l_{i}}\left(N\right)\,\mbox{Tr}_{C}\left[\left(\bgG_{\alpha}c_{m_{1}}^{\alpha}c_{l_{1}}^{\alpha*}\ldots c_{m_{N}}^{\alpha}c_{l_{N}}^{\alpha*}-\bgG_{\beta}c_{m_{1}}^{\beta}c_{l_{1}}^{\beta*}\ldots c_{m_{N}}^{\beta}c_{l_{N}}^{\beta*}\right)\right. \nonumber \\
& & \times\left[\bG^{\text{r}}\left(\omega+\varOmega_{D}\left(p_{1}m_{1}+\ldots+p_{N}m_{N}\right)\right)\left(\bgG-\im\left(p_{1}\left(m_{1}-l_{1}\right)+\ldots+p_{N}\left(m_{N}-l_{N}\right)\right)\right)\bG^{\text{a}}\left(\omega+\varOmega_{D}\left(p_{1}l_{1}+\ldots+p_{N}l_{N}\right)\right)\right] \nonumber \\
& & \left.-\left(\bgG_{\alpha}-\bgG_{\beta}\right)\underset{\gamma}{\sum}c_{m_{1}}^{\gamma}c_{l_{1}}^{\gamma*}\ldots c_{m_{N}}^{\gamma}c_{l_{N}}^{\gamma*}\bG^{\text{r}}\left(\omega+\varOmega_{D}\left(p_{1}m_{1}+\ldots+p_{N}m_{N}\right)\right)\bgG_{\gamma}\bG^{\text{a}}\left(\omega+\varOmega_{D}\left(p_{1}l_{1}+\ldots+p_{N}l_{N}\right)\right)\right] . \label{eq:pumpexpand1}
\eeq
Due to the presence of the generalized Kronecker function $\delta_{m_{i}l_{i}}\left(N\right)$,
the second term on the second line of Eq. (\ref{eq:pumpexpand1})
vanishes, and after some cancellations we arrive at the following:
\beq
I_{\alpha\beta}^{\left(\text{pump}\right)} & = & \frac{1}{\pi}\int d\omega f\left(\omega-\mu\right)\sum_{\mathclap{\substack{m_{1},\ldots,m_{N}\\
l_{1},\ldots,l_{N}}}} \ \delta_{m_{i}l_{i}}\left(N\right)\,\mbox{Tr}_{C}\left[\left(c_{m_{1}}^{\alpha}c_{l_{1}}^{\alpha*}\ldots c_{m_{N}}^{\alpha}c_{l_{N}}^{\alpha*}-c_{m_{1}}^{\beta}c_{l_{1}}^{\beta*}\ldots c_{m_{N}}^{\beta}c_{l_{N}}^{\beta*}\right)\right.\nonumber \\
& & \times\left[\bgG_{\alpha}\bG^{\text{r}}\left(\omega+\varOmega_{D}\left(p_{1}m_{1}+\ldots+p_{N}m_{N}\right)\right)\bgG_{\beta}\bG^{\text{a}}\left(\omega+\varOmega_{D}\left(p_{1}l_{1}+\ldots+p_{N}l_{N}\right)\right)\right. \nonumber \\
& & \left.+\bgG_{\beta}\bG^{\text{r}}\left(\omega+\varOmega_{D}\left(p_{1}m_{1}+\ldots+p_{N}m_{N}\right)\right)\bgG_{\alpha}\bG^{\text{a}}\left(\omega+\varOmega_{D}\left(p_{1}l_{1}+\ldots+p_{N}l_{N}\right)\right)\right] \nonumber \\
& & \left.-\left(\bgG_{\alpha}-\bgG_{\beta}\right)\underset{\gamma\neq\alpha,\beta}{\sum}c_{m_{1}}^{\gamma}c_{l_{1}}^{\gamma*}\ldots c_{m_{N}}^{\gamma}c_{l_{N}}^{\gamma*}\bG^{\text{r}}\left(\omega+\varOmega_{D}\left(p_{1}m_{1}+\ldots+p_{N}m_{N}\right)\right)\bgG_{\gamma}\bG^{\text{a}}\left(\omega+\varOmega_{D}\left(p_{1}l_{1}+\ldots+p_{N}l_{N}\right)\right)\right] . \nonumber \\ \label{eq:generalpump}
\eeq
\end{widetext}

We can immediately derive from equation (\ref{eq:generalpump}) some
general rules for different numbers of harmonics. To focus on the
effect of TR symmetry-breaking, we assume that for all terms except
the $j$-th, the coefficients in each lead are TR symmetric. The $j$-th
harmonic in lead $\beta$ is assumed TR symmetric, and therefore real
(by relation (\ref{eq:realcoeff})), whereas the $j$-th harmonic
in lead $\alpha$ is assumed to break the TR symmetry, so it must
be complex. We also assume that the coefficients of the $j$-th harmonic
have the same magnitude in each lead (i.e. that the \emph{only}
difference between harmonics in leads $\alpha$ and $\beta$ is the
broken TR symmetry of the latter), so that they are related by a multiplicative
phase factor only 
\begin{equation}
c_{m_{j}}^{\alpha}=c_{m_{j}}^{\beta}\ex^{\im\phi_{m_{j}}^{\alpha}} . \label{eq:phaserelate}
\end{equation}
Finally we are in a position to prove some general results for different
choices of $N$.

Let us first consider the case ${N=1}$. In the single harmonic case, coefficients of the different leads are
related by a multiplicative phase factor, and so we obtain: 
\beq
& & \underset{m_{1},l_{1}}{\sum}\left(c_{m_{1}}^{\alpha}c_{l_{1}}^{\alpha*}-c_{m_{1}}^{\beta}c_{l_{1}}^{\beta*}\right)\delta_{m_{i}l_{i}}\left(1\right) \nonumber \\
& = &\underset{m_{1}}{\sum}\left(\left|c_{m_{1}}^{\alpha}\right|^{2}-\left|c_{m_{1}}^{\beta}\right|^{2}\right) = 0 .
\eeq
A pump current between leads $\alpha$ and $\beta$ cannot be guaranteed
in this case unless there are more than two electrodes (if there are
just two, then the final term in Eq. (\ref{eq:generalpump}) is zero),
and $\bgG_{\alpha}\neq\bgG_{\beta}$. For almost
all systems of interest, therefore, $N=1$ implies that $I_{\alpha\beta}^{\left(\text{pump}\right)}=0$.

Then, we have a look at the case ${N\neq1}$. If for all $n_{i}$, $c_{n_{i}}^{\alpha}=c_{n_{i}}^{\beta}$ then
there would trivially be a zero pump current. However, the assumption
(\ref{eq:phaserelate}) means that one can write:
\beq
& & \left(c_{m_{1}}^{\alpha}c_{l_{1}}^{\alpha*}\ldots c_{m_{N}}^{\alpha}c_{l_{N}}^{\alpha*}-c_{m_{1}}^{\beta}c_{l_{1}}^{\beta*}\ldots c_{m_{N}}^{\beta}c_{l_{N}}^{\beta*}\right) \nonumber \\
& = & \left(c_{m_{1}}^{\alpha}c_{l_{1}}^{\alpha*}\ldots c_{m_{j-1}}^{\alpha}c_{l_{j-1}}^{\alpha*}c_{m_{j+1}}^{\alpha}c_{l_{j+1}}^{\alpha*}\ldots c_{m_{N}}^{\alpha}c_{l_{N}}^{\alpha*}\right)\nonumber \\
& & \times c_{m_{j}}^{\alpha}c_{l_{j}}^{\alpha*}\left(\ex^{\im\left(\phi_{m_{j}}^{\alpha}-\phi_{l_{j}}^{\alpha}\right)}-1\right) . \label{eq:amprelate}
\eeq
The generalized Kronecker $\delta_{m_{i}l_{i}}\left(N\right)$ does
retain a term in which $m_{j}=l_{j}$, in which case the pump current
vanishes, but it also returns a term which equals $1$ when, for example,
$p_{j}m_{j}=p_{N}l_{N}$ and $p_{j}l_{j}=p_{N}m_{N}$, as illustrated
in the expansion (\ref{eq:kroneckerexpand}) for $N=2$. This term
will be finite, giving a non-zero pump current in general, and the proof of the quantum pump symmetry theorem is complete.

\section{The adiabatic limit}
\label{app:adiabatic}
As mentioned above, we can always choose the $V_{\alpha}^{\left(n\right)}\left(t\right)$ in Eq.~\eqref{eq:VFS} to be sinusoidal (the $N\to\infty$ case), in which case the $c_{m_{i}}^{\alpha}$ can always be replaced with a summation over Bessel functions of the first kind, via the identity 
\be
\ex^{\im x\sin\left(z\right)}=\overset{\infty}{\underset{r=-\infty}{\sum}}J_{r}\left(x\right)\ex^{\im rz}
\label{eq:BesselID}
\ee
In the sinusoidal series arising from the integral in Eq.~\eqref{eq:phaseINT}, the parameter $x$ always has the dimension of an amplitude over a driving frequency, $A_{\alpha}^{\left(n\right)}/\varOmega_{D}$. Therefore to investigate the adiabatic limit $\varOmega_{D}\to 0$ we can use the asymptotic expression for large $x$ (see e.g. Ref.~\onlinecite{Abramowitz1964}) $J_{r}\left(x\right)\sim\sqrt{\frac{2}{\pi x}}\cos\left(x-\frac{1}{2}r\pi-\frac{1}{4}\pi\right)$ to conclude that the limiting value of the $c_{m_{i}}^{\alpha}$ for low frequencies is vanishing, i.e.
\be 
\underset{\varOmega_{D}\to 0}{\lim}c_{m_{i}}^{\alpha}=0 .
\ee
This means that the adiabatic limit of the pump current is also vanishing:
\be
 \underset{\varOmega_{D}\to 0}{\lim}I_{\alpha\beta}^{\left(pump\right)}=0
\ee
This result is consistent with the prediction of~\citet{Yuge2012} for noninteracting systems that there is zero pump current when only the chemical potentials of the leads are adiabatically modulated in time.

\section{Details of the TD and time-averaged calculations}\label{app:details}
We can remove all frequency integrals in Eq.~\eqref{eq: CURRENT} after expanding the Fermi function into a series expansion whose terms
possess a simple pole structure~\cite{Ozaki2007}:
\begin{equation}
f\left(x\right)=\frac{1}{\ex^{\beta x}+1}=\frac{1}{2}-\underset{N_{p}\to \infty}{\lim}\underset{l=1}{\overset{N_{p}}{\sum}}\eta_{l}\left(\frac{1}{\beta x+\im\zeta_{l}}+\frac{1}{\beta x-\im\zeta_{l}}\right) . \label{eq:PADE}
\end{equation}
When the parameter values are $\eta_{l}=1$ and $\zeta_{l}=\pi\left(2l-1\right)$,
this is referred to as the Matsubara expansion, but one can also improve
the convergence of this series for finite $N_{p}$ by expressing the poles of the
Fermi function as the solutions to an eigenproblem for
a tridiagonal matrix \cite{Croy2009,Hu2010,Hu2011}, in the so-called Pad{\'e} expansion.

To deal numerically with formulas containing the effective Hamiltonian $\bH_{CC}^{\text{\text{eff}}}$, we introduce the left and right eigenvectors:
\begin{eqnarray}
\bH_{CC}^{\text{eff}}\left|\varphi_{j}^{\text{R}}\right\rangle  & = & \bar{\varepsilon}_{j}\left|\varphi_{j}^{\text{R}}\right\rangle \,\,\mbox{and}\,\,\left\langle \varphi_{j}^{\text{L}}\right|\bH_{CC}^{\text{eff}}=\bar{\varepsilon}_{j}\left\langle \varphi_{j}^{\text{L}}\right| . \label{eq:leftrighteigenproblem}
\end{eqnarray}
All expressions containing the effective Hamiltonian can be recast into summations over left/right eigenvectors using the following idempotency relation:
\begin{equation}
\underset{j}{\sum}\frac{\left|\varphi_{j}^{\text{R}}\right\rangle \left\langle \varphi_{j}^{\text{L}}\right|}{\left\langle \varphi_{j}^{\text{L}}\mid\varphi_{j}^{\text{R}}\right\rangle }=\unit=\underset{j}{\sum}\frac{\left|\varphi_{j}^{\text{L}}\right\rangle \left\langle \varphi_{j}^{\text{R}}\right|}{\left\langle \varphi_{j}^{\text{R}}\mid\varphi_{j}^{\text{L}}\right\rangle } . \label{eq:idempotency}
\end{equation}
By inserting the expansion for the Fermi function in Eq. \eqref{eq: CURRENT} and removing all
frequency integrals, it is possible to evaluate exactly the time-dependent current
in terms of a set of special functions. The first is the so-called \emph{Hurwitz-Lerch Transcendent} $\F$~\cite{Lerch1887}:
\begin{equation}
\F\left(z,s,a\right)\equiv\underset{n=0}{\overset{\infty}{\sum}}\frac{z^{n}}{\left(n+a\right)^{s}} . \label{eq:HLT}
\end{equation}
We will also use the \emph{Digamma function}, defined as the logarithmic derivative of the gamma function, $\Q(z)=d\Gamma(z)/dz$. The functions $\F$ and $\Q$ appear after we use the Matsubara parameters in Eq. \eqref{eq:PADE} and identify these special functions with the resulting infinite summations. They can be evaluated to arbitrary accuracy either by using an equivalent Pad{\'e} expansion or by using known numerical routines~\cite{hypf-impl,gsl,Tuovinen2016b}. We also define the following object as a series expansion:
\beq
& & \mbox{cosech}\left.\left(\frac{\pi}{\beta}\left(t_{1}-t_{2}\right)\right)\right|_{t_{1}\neq t_{2}} \nonumber \\
& \simeq & 2\underset{l=1}{\overset{N_{p}}{\sum}}\eta_{l}\left[\theta\left(t_{1}-t_{2}\right)\ex^{-\frac{\zeta_{l}}{\beta}\left(t_{1}-t_{2}\right)}\right.\nonumber\\
& & \left.\hspace{30pt}- \ \theta\left(t_{2}-t_{1}\right)\ex^{-\frac{\zeta_{l}}{\beta}\left(t_{2}-t_{1}\right)}\right]\label{eq:cosech}
\eeq
where this is set to zero when $t_{1}=t_{2}$. For an arbitrary time-dependent bias, the function
$\mbox{cosech}\left.\left(\frac{\pi}{\beta}\left(t_{1}-t_{2}\right)\right)\right|_{t_{1}\neq t_{2}}$ may be implemented using either the Matsubara parameters or the Pad{\'e} parameters as in Refs.~\cite{Ridley2016a, Ridley2017}. We also note that the step function $\theta$ is defined by the midpoint convention. With these definitions, it is possible to expand the time-dependent current for an \emph{arbitrary} time-dependent bias thus:
\begin{widetext}
\beq
I_{\alpha}\left(t\right) & = & \frac{1}{\pi}\underset{j}{\sum}\left\{ \mbox{Re}\left[2\frac{\left\langle \varphi_{j}^{\text{L}}\right|\bgG_{\alpha}\left|\varphi_{j}^{\text{R}}\right\rangle }{\left\langle \varphi_{j}^{\text{L}}\mid\varphi_{j}^{\text{R}}\right\rangle }\left( \phantom{\int_{t_0}^t} \right.\right.\right. \nonumber \\
& & \left.\left.\left.-\im\ex^{\im\psi_{\alpha}\left(t,t_{0}\right)}\ex^{-\im\left(\bar{\varepsilon}_{j}-\mu\right)\left(t-t_{0}\right)}\bar{\F}\left(t-t_{0},\beta,-\left(\bar{\varepsilon}_{j}-\mu\right)\right) - \frac{\im\pi}{\beta}\int_{t_0}^t d\tau \ex^{-\im\left(\bar{\varepsilon}_{j}-\mu\right)\left(t-\tau\right)}\ex^{\im\psi_{\alpha}\left(t,\tau\right)}\left.\mbox{cosech}\left(\frac{\pi}{\beta}\left(t-\tau\right)\right)\right|_{t\neq\tau}\right)\right]\right.\nonumber \\
& & \left.-\underset{\gamma,k}{\sum}\frac{\left\langle \varphi_{k}^{\text{R}}\right|\bgG_{\alpha}\left|\varphi_{j}^{\text{R}}\right\rangle \left\langle \varphi_{j}^{\text{L}}\right|\bgG_{\gamma}\left|\varphi_{k}^{\text{L}}\right\rangle }{\left\langle \varphi_{j}^{\text{L}}\mid\varphi_{j}^{\text{R}}\right\rangle \left\langle \varphi_{k}^{\text{R}}\mid\varphi_{k}^{\text{L}}\right\rangle }\ex^{-\im\left(\bar{\varepsilon}_{j}-\bar{\varepsilon}_{k}^{*}\right)\left(t-t_{0}\right)}\left[\frac{\Psi\left(\frac{1}{2}+\frac{\beta}{2\pi\im}\left(\bar{\varepsilon}_{k}^{*}-\mu\right)\right)-\Psi\left(\frac{1}{2}-\frac{\beta}{2\pi\im}\left(\bar{\varepsilon}_{j}-\mu\right)\right)}{\bar{\varepsilon}_{k}^{*}-\bar{\varepsilon}_{j}}\right.\right. \nonumber \\
& & \left.\left.+\int_{t_0}^t d\bar{t}\left[-\im\ex^{-\im\left(\bar{\varepsilon}_{k}^{*}-\mu\right)\left(\bar{t}-t_{0}\right)}\ex^{\im\psi_{\beta}\left(\bar{t},t_{0}\right)}\bar{\F}\left(\bar{t}-t_{0},\beta,-\left(\bar{\varepsilon}_{j}-\mu\right)\right)+\text{c.c.}_{j\leftrightarrow k}\right]+\frac{2\pi\im}{\beta}I_{\gamma}\left(t,\beta,\mu,\bar{\varepsilon}_{j},\bar{\varepsilon}_{k}^{*}\right)\right]\right\} . \label{eq:CURRENTFINAL}
\eeq
Here, we have defined the function:
\begin{equation}
I_{\gamma}\left(t,\beta,\mu,\bar{\varepsilon}_{j},\bar{\varepsilon}_{k}^{*}\right)=\frac{1}{2}\int_{t_0}^t d\tau\int_{t_0}^t d\bar{\tau}\ex^{\im\left(\bar{\varepsilon}_{j}-\mu\right)\left(\tau-t_{0}\right)}\ex^{-\im\left(\bar{\varepsilon}_{k}^{*}-\mu\right)\left(\bar{\tau}-t_{0}\right)}\ex^{-\im\psi_{\gamma}\left(\tau,\bar{\tau}\right)}\left.\mbox{cosech}\left(\frac{\pi}{\beta}\left(\tau-\bar{\tau}\right)\right)\right|_{\tau\neq\bar{\tau}} . \label{eq:Ifunction}
\end{equation}
We have also introduced the following compact notation for terms involving the
Hurwitz-Lerch Transcendent: 
\begin{equation}
\bar{\F}\left(\beta,\tau,z\right)\equiv\exp\left(-\frac{\pi}{\beta}\tau\right)\F\left(\ex^{-\frac{2\pi\tau}{\beta}},1,\frac{1}{2}+\frac{\beta z}{2\pi\im}\right) . \label{eq:phigen-1}
\end{equation}

This general result, Eq.~\eqref{eq:CURRENTFINAL} is valid for all time-dependent biases. However, for the purposes of this paper we may substitute the biharmonic bias Eq.~\eqref{eq:biharmonicbias} into Eq.~\eqref{eq:CURRENTFINAL} by expanding the exponential phase factor in terms of Bessel functions of the first kind via Eq. \eqref{eq:BesselID}:
\beq
& & \ex^{\im\psi_{\alpha}\left(t_{1},t_{2}\right)} = \ex^{\im V_{\alpha}\left(t_{1}-t_{2}\right)}\underset{r,r',s,s'}{\sum}J_{r}\left(\frac{A_{\alpha}^{\left(1\right)}}{p_{1}\varOmega_{\alpha}}\right)J_{r'}\left(\frac{A_{\alpha}^{\left(1\right)}}{p_{1}\varOmega_{\alpha}}\right) J_{s}\left(\frac{A_{\alpha}^{\left(2\right)}}{p_{2}\varOmega_{\alpha}}\right)J_{s'}\left(\frac{A_{\alpha}^{\left(2\right)}}{p_{2}\varOmega_{\alpha}}\right) \nonumber \\
& & \times e^{i\left(r-r'\right)\phi_{\alpha}}e^{i\Omega_{\alpha}\left(p_{1}r+p_{2}s\right)\left(t_{1}-t_{0}\right)}e^{-i\Omega_{\alpha}\left(p_{1}r'+p_{2}s'\right)\left(t_{2}-t_{0}\right)}. 
\eeq
We note in passing that this has already been done for the current fluctuations in Ref.~\onlinecite{Ridley2017}, and using the same logic we can remove all time integrals appearing in Eq.~\eqref{eq:CURRENTFINAL}:
\beq
& & I_{\alpha}\left(t\right) = \frac{1}{\pi}\underset{j}{\sum}\left\{ 2\textrm{Re}\left[\frac{\left\langle \varphi_{j}^{\text{L}}\right|\bgG_{\alpha}\left|\varphi_{j}^{\text{R}}\right\rangle }{\left\langle \varphi_{j}^{\text{L}}\mid\varphi_{j}^{\text{R}}\right\rangle }\left[\phantom{\left(\frac{A_{\alpha}^{\left(2\right)}}{p_{2}\varOmega_{\alpha}}\right)}\right.\right.\right.\nonumber \\
& & \left.\left.\left.\im\ex^{\im\frac{A_{\alpha}^{\left(1\right)}}{p_{1}\varOmega_{\alpha}}\left(\sin\left(p_{1}\varOmega_{\alpha}\left(t-t_{0}\right)+\phi_{\alpha}\right)-\sin\phi_{\alpha}\right)}\ex^{\im\frac{A_{\alpha}^{\left(2\right)}}{p_{2}\varOmega_{\alpha}}\sin\left(p_{2}\varOmega_{\alpha}\left(t-t_{0}\right)\right)}\underset{r,s}{\sum}J_{r}\left(\frac{A_{\alpha}^{\left(1\right)}}{p_{1}\varOmega_{\alpha}}\right)J_{s}\left(\frac{A_{\alpha}^{\left(2\right)}}{p_{2}\varOmega_{\alpha}}\right)\ex^{-\im r\phi_{\alpha}}\right.\right.\right.\nonumber \\
& & \left.\left.\left.\times\left[\ex^{-\im\left(\bar{\varepsilon}_{j}-\mu-V_{\alpha}\right)\left(t-t_{0}\right)}\left[\bar{\F}\left(t-t_{0},\beta,-\left(\bar{\varepsilon}_{j}-\mu-V_{\alpha}-\varOmega_{\alpha}\left(p_{1}r+p_{2}s\right)\right)\right)-\bar{\F}\left(t-t_{0},\beta,-\left(\bar{\varepsilon}_{j}-\mu\right)\right)\right]\right.\right.\right.\right. \nonumber \\
& & \left.\left.\left.\left.+\ex^{-\im\varOmega_{\alpha}\left(p_{1}r+p_{2}s\right)\left(t-t_{0}\right)}\Psi\left(\frac{1}{2}-\frac{\beta}{2\pi \im}\left(\bar{\varepsilon}_{j}-\mu-V_{\alpha}-\varOmega_{\alpha}\left(p_{1}r+p_{2}s\right)\right)\right)\right]\right]\right]\right. \nonumber \\ \nonumber \\ \nonumber \\
& & \left.-\underset{\gamma,k}{\sum}\frac{\left\langle \varphi_{k}^{\text{R}}\right|\bgG_{\alpha}\left|\varphi_{j}^{\text{R}}\right\rangle \left\langle \varphi_{j}^{\text{L}}\right|\bgG_{\gamma}\left|\varphi_{k}^{\text{L}}\right\rangle }{\left\langle \varphi_{j}^{\text{L}}\mid\varphi_{j}^{\text{R}}\right\rangle \left\langle \varphi_{k}^{\text{R}}\mid\varphi_{k}^{\text{L}}\right\rangle }\left[\frac{\ex^{-\im\left(\bar{\varepsilon}_{j}-\bar{\varepsilon}_{k}^{*}\right)\left(t-t_{0}\right)}}{\bar{\varepsilon}_{k}^{*}-\bar{\varepsilon}_{j}}\left[\Psi\left(\frac{1}{2}+\frac{\beta}{2\pi \im}\left(\bar{\varepsilon}_{k}^{*}-\mu\right)\right)-\Psi\left(\frac{1}{2}-\frac{\beta}{2\pi \im}\left(\bar{\varepsilon}_{j}-\mu\right)\right)\right]\right.\right. \nonumber \\
& & \left.\left.+\underset{r,s}{\sum}J_{r}\left(\frac{A_{\gamma}^{\left(1\right)}}{p_{1}\varOmega_{\gamma}}\right)J_{s}\left(\frac{A_{\gamma}^{\left(2\right)}}{p_{2}\varOmega_{\gamma}}\right)\left[\frac{\ex^{-\im r\phi_{\gamma}}\ex^{\im\frac{A_{\gamma}^{\left(1\right)}}{p_{1}\varOmega_{\gamma}}\sin\phi_{\gamma}}}{\bar{\varepsilon}_{j}-\bar{\varepsilon}_{k}^{*}-V_{\gamma}-\varOmega_{\gamma}\left(p_{1}r+p_{2}s\right)}\right.\right.\right. \nonumber \\
& & \left.\left.\left.\times\left[\ex^{-\im\left(\bar{\varepsilon}_{j}-\bar{\varepsilon}_{k}^{*}\right)\left(t-t_{0}\right)}\left[\Psi\left(\frac{1}{2}+\frac{\beta}{2\pi \im}\left(\bar{\varepsilon}_{k}^{*}-\mu\right)\right)-\Psi\left(\frac{1}{2}+\frac{\beta}{2\pi \im}\left(\bar{\varepsilon}_{j}-\mu-V_{\gamma}-\varOmega_{\gamma}\left(p_{1}r+p_{2}s\right)\right)\right)\right]\right.\right.\right.\right.\nonumber \\
& & \left.\left.\left.\left.+\ex^{\im\left(\bar{\varepsilon}_{k}^{*}-\mu-V_{\gamma}-\varOmega_{\gamma}\left(p_{1}r+p_{2}s\right)\right)\left(t-t_{0}\right)}\left[\bar{\F}\left(t-t_{0},\beta,\bar{\varepsilon}_{k}^{*}-\mu\right)-\bar{\F}\left(t-t_{0},\beta,\bar{\varepsilon}_{j}-\mu-V_{\gamma}-\varOmega_{\gamma}\left(p_{1}r+p_{2}s\right)\right)\right]\right]\right.\right.\right.\nonumber \\
& & \left.\left.\left.+\frac{\ex^{\im r\phi_{\gamma}}\ex^{-\im\frac{A_{\gamma}^{\left(1\right)}}{p_{1}\varOmega_{\gamma}}\sin\phi_{\gamma}}}{\bar{\varepsilon}_{k}^{*}-\bar{\varepsilon}_{j}-V_{\gamma}-\varOmega_{\gamma}\left(p_{1}r+p_{2}s\right)}\right.\right.\right.\nonumber \\
& & \left.\left.\left.\times\left[\ex^{-\im\left(\bar{\varepsilon}_{j}-\bar{\varepsilon}_{k}^{*}\right)\left(t-t_{0}\right)}\left[\Psi\left(\frac{1}{2}-\frac{\beta}{2\pi \im}\left(\bar{\varepsilon}_{j}-\mu\right)\right)-\Psi\left(\frac{1}{2}-\frac{\beta}{2\pi \im}\left(\bar{\varepsilon}_{k}^{*}-\mu-V_{\gamma}-\varOmega_{\gamma}\left(p_{1}r+p_{2}s\right)\right)\right)\right]\right.\right.\right.\right.\nonumber \\
& & \left.\left.\left.\left.+\ex^{-\im\left(\bar{\varepsilon}_{j}-\mu-V_{\gamma}-\varOmega_{\gamma}\left(p_{1}r+p_{2}s\right)\right)\left(t-t_{0}\right)}\left[\bar{\F}\left(t-t_{0},\beta,-\left(\bar{\varepsilon}_{j}-\mu\right)\right)-\bar{\F}\left(t-t_{0},\beta,-\left(\bar{\varepsilon}_{k}^{*}-\mu-V_{\gamma}-\varOmega_{\gamma}\left(p_{1}r+p_{2}s\right)\right)\right)\right]\right]\right]\right.\right.\nonumber \\ \nonumber \\ \nonumber \\
& & \left.\left.+\underset{r,r',s,s'}{\sum}J_{r}\left(\frac{A_{\gamma}^{\left(1\right)}}{p_{1}\varOmega_{\gamma}}\right)J_{r'}\left(\frac{A_{\gamma}^{\left(1\right)}}{p_{1}\varOmega_{\gamma}}\right)J_{s}\left(\frac{A_{\gamma}^{\left(2\right)}}{p_{2}\varOmega_{\gamma}}\right)J_{s'}\left(\frac{A_{\gamma}^{\left(2\right)}}{p_{2}\varOmega_{\gamma}}\right)\frac{\ex^{-\im\left(r-r'\right)\phi_{\gamma}}}{\bar{\varepsilon}_{j}-\bar{\varepsilon}_{k}^{*}-\varOmega_{\gamma}\left(p_{1}\left(r-r'\right)+p_{2}(s-s')\right)}\right.\right.\nonumber \\
& & \left.\left.\times\left[\ex^{-\im\varOmega_{\gamma}\left(p_{1}\left(r-r'\right)+p_{2}\left(s-s'\right)\right)\left(t-t_{0}\right)}\right.\right.\right.\nonumber \\
& & \left.\left.\left.\times\left[\Psi\left(\frac{1}{2}-\frac{\beta}{2\pi \im}\left(\bar{\varepsilon}_{j}-\mu-V_{\gamma}-\varOmega_{\gamma}\left(p_{1}r+p_{2}s\right)\right)\right)-\Psi\left(\frac{1}{2}+\frac{\beta}{2\pi \im}\left(\bar{\varepsilon}_{k}^{*}-\mu-V_{\gamma}-\varOmega_{\gamma}\left(p_{1}r'+p_{2}s'\right)\right)\right)\right]\right.\right.\right.\nonumber \\
& & \left.\left.\left.+\ex^{-\im\left(\bar{\varepsilon}_{j}-\bar{\varepsilon}_{k}^{*}\right)\left(t-t_{0}\right)}\left[\Psi\left(\frac{1}{2}+\frac{\beta}{2\pi \im}\left(\bar{\varepsilon}_{j}-\mu-V_{\gamma}-\varOmega_{\gamma}\left(p_{1}r+p_{2}s\right)\right)\right)-\Psi\left(\frac{1}{2}-\frac{\beta}{2\pi \im}\left(\bar{\varepsilon}_{k}^{*}-\mu-V_{\gamma}-\varOmega_{\gamma}\left(p_{1}r'+p_{2}s'\right)\right)\right)\right]\right.\right.\right. \nonumber \\ \nonumber \\
& & \left.\left.\left.+\ex^{\im \left(\bar{\varepsilon}_{k}^{*}-\mu-V_{\gamma}-\varOmega_{\gamma}\left(p_{1}r+p_{2}s\right)\right)\left(t-t_{0}\right)}\right.\right.\right.\nonumber \\
& & \left.\left.\left.\times\left[\bar{\F}\left(t-t_{0},\beta,\bar{\varepsilon}_{j}-\mu-V_{\gamma}-\varOmega_{\gamma}\left(p_{1}r+p_{2}s\right)\right)-\bar{\F}\left(t-t_{0},\beta,\bar{\varepsilon}_{k}^{*}-\mu-V_{\gamma}-\varOmega_{\gamma}\left(p_{1}r'+p_{2}s'\right)\right)\right]\right.\right.\right. \nonumber \\ \nonumber \\
& & \left.\left.\left.+\ex^{-\im\left(\bar{\varepsilon}_{j}-\mu-V_{\gamma}-\varOmega_{\gamma}\left(p_{1}r'+p_{2}s'\right)\right)\left(t-t_{0}\right)}\right.\right.\right. \nonumber \\
& & \left.\left.\left.\times\left[\bar{\F}\left(t-t_{0},\beta,-\left(\bar{\varepsilon}_{j}-\mu-V_{\gamma}-\varOmega_{\gamma}\left(p_{1}r+p_{2}s\right)\right)\right)-\bar{\F}\left(t-t_{0},\beta,-\left(\bar{\varepsilon}_{k}^{*}-\mu-V_{\gamma}-\varOmega_{\gamma}\left(p_{1}r'+p_{2}s'\right)\right)\right)\right]\right]\right]\right\} . \label{eq:CURRENTFINALBESS}
\eeq
We note that the above formula is obtained by using the Matsubara parameters in Eq.~\eqref{eq:cosech} and identifying all resulting infinite expansions with either the $\bar{\F}$ or $\Q$ functions defined above. It is therefore an exact expression, and is the expression used to obtain our numerical results for the time-dependent current in the different GNR configurations.

We now assume that the fundamental driving frequency is lead-independent,
$\varOmega_{\gamma}=\varOmega_{D}$, for all $\gamma$. This allows us to
derive a formula for the time-independent, steady-state pump current
using the definitions in Eqs. (\ref{eq:pump}) and  (\ref{eq:stablepump}):
\beq
I_{\alpha\beta}^{\left(\text{pump}\right)} & = & \frac{1}{\pi}\underset{\gamma}{\sum}\underset{j,k}{\sum}\underset{r,r',s,s'}{\sum}\delta_{ss}^{rr'}\left(p_{i}\right)\left[\frac{\left\langle \varphi_{k}^{\text{R}}\right|\bgG_{\alpha}\left|\varphi_{j}^{\text{R}}\right\rangle \left\langle \varphi_{j}^{\text{L}}\right|\bgG_{\gamma}\left|\varphi_{k}^{\text{L}}\right\rangle }{\left\langle \varphi_{j}^{\text{L}}\mid\varphi_{j}^{\text{R}}\right\rangle \left\langle \varphi_{k}^{\text{R}}\mid\varphi_{k}^{\text{L}}\right\rangle \left(\bar{\varepsilon}_{j}-\bar{\varepsilon}_{k}^{*}\right)}\left[ \phantom{\left(\frac{A_{\alpha}^{\left(2\right)}}{p_{2}\varOmega_{D}}\right)} \right.\right.\nonumber \\
& & \left.\left. \ex^{-\im\left(r-r'\right)\phi_{\alpha}}J_{r}\left(\frac{A_{\alpha}^{\left(1\right)}}{p_{1}\varOmega_{D}}\right)J_{r'}\left(\frac{A_{\alpha}^{\left(1\right)}}{p_{1}\varOmega_{D}}\right)J_{s}\left(\frac{A_{\alpha}^{\left(2\right)}}{p_{2}\varOmega_{D}}\right)J_{s'}\left(\frac{A_{\alpha}^{\left(2\right)}}{p_{2}\varOmega_{D}}\right)\right.\right. \nonumber \\
& & \left.\left.\times\left[\Psi\left(\frac{1}{2}-\frac{\beta}{2\pi \im}\left(\bar{\varepsilon}_{j}-\mu-V_{\alpha}-\varOmega_{D}\left(p_{1}r+p_{2}s\right)\right)\right)-\Psi\left(\frac{1}{2}+\frac{\beta}{2\pi \im}\left(\bar{\varepsilon}_{k}^{*}-\mu-V_{\alpha}-\varOmega_{D}\left(p_{1}r'+p_{2}s'\right)\right)\right)\right] \right.\right. \nonumber \\ \nonumber \\
& & \left.\left.-\ex^{-\im\left(r-r'\right)\phi_{\gamma}}J_{r}\left(\frac{A_{\gamma}^{\left(1\right)}}{p_{1}\varOmega_{D}}\right)J_{r'}\left(\frac{A_{\gamma}^{\left(1\right)}}{p_{1}\varOmega_{D}}\right)J_{s}\left(\frac{A_{\gamma}^{\left(2\right)}}{p_{2}\varOmega_{D}}\right)J_{s'}\left(\frac{A_{\gamma}^{\left(2\right)}}{p_{2}\varOmega_{D}}\right)\right.\right.\nonumber \\
& & \left.\left.\left.\times\left[\Psi\left(\frac{1}{2}-\frac{\beta}{2\pi \im}\left(\bar{\varepsilon}_{j}-\mu-V_{\gamma}-\varOmega_{D}\left(p_{1}r+p_{2}s\right)\right)\right)-\Psi\left(\frac{1}{2}+\frac{\beta}{2\pi \im}\left(\bar{\varepsilon}_{k}^{*}-\mu-V_{\gamma}-\varOmega_{D}\left(p_{1}r'+p_{2}s'\right)\right)\right)\right]\right] \right.\right. \nonumber \\ \nonumber \\ \nonumber \\
& & \left.\left.-\frac{\left\langle \varphi_{k}^{\text{R}}\right|\bgG_{\beta}\left|\varphi_{j}^{\text{R}}\right\rangle \left\langle \varphi_{j}^{\text{L}}\right|\bgG_{\gamma}\left|\varphi_{k}^{\text{L}}\right\rangle }{\left\langle \varphi_{j}^{\text{L}}\mid\varphi_{j}^{\text{R}}\right\rangle \left\langle \varphi_{k}^{\text{R}}\mid\varphi_{k}^{\text{L}}\right\rangle \left(\bar{\varepsilon}_{j}-\bar{\varepsilon}_{k}^{*}\right)}\left[\ex^{-\im\left(r-r'\right)\phi_{\beta}}J_{r}\left(\frac{A_{\beta}^{\left(1\right)}}{p_{1}\varOmega_{D}}\right)J_{r'}\left(\frac{A_{\beta}^{\left(1\right)}}{p_{1}\varOmega_{D}}\right)J_{s}\left(\frac{A_{\beta}^{\left(2\right)}}{p_{2}\varOmega_{D}}\right)J_{s'}\left(\frac{A_{\beta}^{\left(2\right)}}{p_{2}\varOmega_{D}}\right)\right.\right.\right. \nonumber \\
& & \left.\left.\left.\times\left[\Psi\left(\frac{1}{2}-\frac{\beta}{2\pi \im}\left(\bar{\varepsilon}_{j}-\mu-V_{\beta}-\varOmega_{D}\left(p_{1}r+p_{2}s\right)\right)\right)-\Psi\left(\frac{1}{2}+\frac{\beta}{2\pi \im}\left(\bar{\varepsilon}_{k}^{*}-\mu-V_{\beta}-\varOmega_{D}\left(p_{1}r'+p_{2}s'\right)\right)\right)\right] \right.\right.\right.\nonumber \\ \nonumber \\
& & \left.\left.\left.-\ex^{-\im\left(r-r'\right)\phi_{\gamma}}J_{r}\left(\frac{A_{\gamma}^{\left(1\right)}}{p_{1}\varOmega_{D}}\right)J_{r'}\left(\frac{A_{\gamma}^{\left(1\right)}}{p_{1}\varOmega_{D}}\right)J_{s}\left(\frac{A_{\gamma}^{\left(2\right)}}{p_{2}\varOmega_{D}}\right)J_{s'}\left(\frac{A_{\gamma}^{\left(2\right)}}{p_{2}\varOmega_{D}}\right) \right.\right.\right. \nonumber \\
& & \left.\left.\times\left[\Psi\left(\frac{1}{2}-\frac{\beta}{2\pi \im}\left(\bar{\varepsilon}_{j}-\mu-V_{\gamma}-\varOmega_{D}\left(p_{1}r+p_{2}s\right)\right)\right)-\Psi\left(\frac{1}{2}+\frac{\beta}{2\pi \im}\left(\bar{\varepsilon}_{k}^{*}-\mu-V_{\gamma}-\varOmega_{D}\left(p_{1}r'+p_{2}s'\right)\right)\right)\right]\right]\right] . \nonumber \\ \label{eq:pumparbitrary}
\eeq
In this expression, we have used the modified Kronecker function
\begin{equation}
\delta_{ss'}^{rr'}\left(p_{i}\right)\equiv\begin{cases}
\begin{array}{c}
1,\, p_{1}\left(r-r'\right)+p_{2}\left(s-s'\right)=0\\
0,\,\textrm{else}
\end{array}\end{cases}\label{eq:Kronp1p2}
\end{equation}
to make the replacement $\bar{\varepsilon}_{j}-\bar{\varepsilon}_{k}^{*}-\varOmega_{D}\left(p_{1}\left(r-r'\right)+p_{2}(s-s')\right)\leftrightarrow\bar{\varepsilon}_{j}-\bar{\varepsilon}_{k}^{*}$. Eq.~\eqref{eq:pumparbitrary} forms the basis for the static quantum pump calculations carried out in Section \ref{sec:results}.
\end{widetext}


%


\end{document}